\documentclass[12pt]{article}
\usepackage[paperheight=11in,paperwidth=8.5in]{geometry}
\usepackage{amsthm,amsmath,latexsym,multibox,amssymb,amsfonts,amscd}
\usepackage{graphicx,lscape,fancyhdr,array,stmaryrd,euscript,wrapfig}
\newcolumntype{x}[1]{%
>{\centering\hspace{0pt}}m{#1}}%

\newcommand{\cc}{\mathcal{C}}
\newcommand{\p}[1]{|#1|}

\newcommand{\dl}[1]{\mathchoice{\ffrac{\dd}{\dd #1}}{\frac{\dd}{\dd
      #1}}{\ffrac{\dd}{\dd #1}}{\ffrac{\dd}{\dd #1}}}

\newcommand{\ddl}[2]{\ffrac{\dd #1}{\dd #2}}

\newcommand{\ffrac}[2]{\raisebox{.5pt}%
  {\footnotesize$\displaystyle\frac{#1}{#2}$}\kern1pt}
  \newcommand{\gh}[1]{\mathrm{gh}(#1)}
  
\newcommand{\manifold}[1]{\mathcal{#1}}

\newcommand{\manM}{\manifold{M}}

\newcommand{\Liealg}{\mathfrak}
\newcommand{\algg}{\Liealg{g}}

\newcommand{\algh}{\Liealg{h}}

\newcommand{\algA}{\mathfrak{A}}
\newcommand{\algB}{\mathfrak{hs}}

\def\cF{\mathcal{F}}

\newcommand{\commut}[2]{[#1{,}\,#2]}
\newcommand{\scommut}[2]{\{#1{,}\,#2\}}
\newcommand{\qcommut}[2]{[#1{,}\,#2]_\star}
\newcommand{\bref}[1]{\textbf{\ref{#1}}}

\newcommand{\half}{\mathchoice{%
    \ffrac{1}{2}}{\frac{1}{2}}{\frac{1}{2}}{\frac{1}{2}}}

\newcommand{\pl}{\partial}
\newcommand{\dd}{\partial}

\newcommand{\be}{\begin{equation}}
\newcommand{\ee}{\end{equation}}
\newcommand{\bes}{\begin{split}}
\newcommand{\es}{\end{split}}
\newcommand{\bee}{\begin{eqnarray}}
\newcommand{\eee}{\end{eqnarray}}
\newcommand{\beee}{\begin{array}}
\newcommand{\bem}{\begin{multline}}
\newcommand{\eem}{\end{multline}}
\newcommand{\bec}{\begin{Comment}}
\newcommand{\ec}{\end{Comment}}

\newcommand{\mm}{{\ensuremath{\underline{m}}}}

\newcommand{\aAs}{{\ensuremath{\mathsf{A}}}}
\newcommand{\aBs}{{\ensuremath{\mathsf{B}}}}

\newcommand{\aA}{{\ensuremath{\mathcal{A}}}}
\newcommand{\aB}{{\ensuremath{\mathcal{B}}}}
\newcommand{\aC}{{\ensuremath{\mathcal{C}}}}
\newcommand{\aD}{{\ensuremath{\mathcal{D}}}}

\newcommand{\gad}{{\dot{\alpha}}}
\newcommand{\gbd}{{\dot{\beta}}}

\newcommand{\ga}{\alpha}
\newcommand{\gb}{\beta}
\newcommand{\gc}{\gamma}
\newcommand{\gd}{\delta}

\newcommand{\fud}[2]{{}^{#1}{}_{#2}\,}

\newcommand{\besubeqs}{\begin{subequations}}
\newcommand{\esubeqs}{\end{subequations}}

\usepackage{cite}

\usepackage{rotating,framed}
\newcommand{\subsB}[2]{{\substack{\displaystyle#1\\ \\{\displaystyle#2}}}}
\newcommand{\subsC}[3]{{\substack{\displaystyle#1\\ \\{\displaystyle#2}\\ \\{\displaystyle#3}}}}

\usepackage{hyperref}
\numberwithin{equation}{section} \makeatletter
\@addtoreset{equation}{section}

\usepackage[toc]{multitoc}

\begin{document}
\begin{flushright}
FIAN-TD-2014-15 \\
AEI-2014-045
\end{flushright}

\vspace{1cm}

\begin{center}

{\Large\textbf{Uniformizing higher-spin equations}}

\vspace{.9cm}

{\large K.B. Alkalaev$^{\,a,b}$,  M.A. Grigoriev$^{\,a,b}$, and E.D. Skvortsov$^{\,c,a}$}

\vspace{0.5cm}

\textit{$^{a}$I.E. Tamm Department of Theoretical Physics, \\P.N. Lebedev Physical
Institute,\\ Leninsky ave. 53, 119991 Moscow, Russia}

\vspace{0.5cm}

\textit{$^{b}$Moscow Institute of Physics and Technology, Dolgoprudnyi,\\
141700 Moscow region, Russia}

\vspace{0.5cm}

\textit{$^c$ Albert Einstein Institute, Golm, Germany, D-14476, Am M\"{u}hlenberg 1}

\vspace{0.5cm}

\thispagestyle{empty}


\end{center}

\begin{abstract}

Vasiliev's higher-spin theories in various dimensions  are uniformly represented as a simple system of equations. These equations and
their gauge invariances are based on two superalgebras and have a transparent algebraic meaning.
For a given higher-spin theory these algebras can be inferred from  the vacuum   higher-spin symmetries. The proposed system of equations admits a concise AKSZ formulation.  We also discuss novel higher-spin  systems including partially-massless and massive fields in AdS, as well as conformal and massless  off-shell fields.

\end{abstract}
{ \footnotesize
\tableofcontents }

\newpage
\section{Introduction}

In this paper we study algebraic structures
underlying Vasiliev's higher-spin (HS) theories in various
dimensions~\cite{Vasiliev:1990en,Vasiliev:1992av, Vasiliev:1992ix,
Prokushkin:1998vn, Vasiliev:2003ev} (see \cite{Vasiliev:1999ba,
Bekaert:2005vh,Didenko:2014dwa} for reviews). These HS theories are defined by classical equations of motion whose
underlying geometrical principles are not entirely settled. More precisely,
the equations of interacting higher-spin fields are encoded~\cite{Vasiliev:1990vu} in the flatness condition imposed on the connection of a sufficiently large algebra, called the {\em embedding algebra} in this work. The connection is further constrained through its coupling to a set of 0-form fields which in turn are subject to certain algebraic constraints.

Another algebraic ingredient of the HS theory is an algebra of
{\em dynamical symmetries} which is somewhat hidden in the usual
formulations. This is a finite-dimensional Lie
superalgebra that remains undeformed at the interacting level. On the other hand the infinite-dimensional {\em higher-spin  algebra} is a subalgebra of
the {\em embedding algebra} preserving the most symmetric vacuum,
which corresponds to empty $AdS$ space. We argue that in HS theories
the embedding algebra can be constructed as a twisted
tensor product of the higher-spin  algebra and the algebra of
dynamical symmetries. In particular, both factors turn
out to be subalgebras of the embedding  algebra.

In this paper we propose a formulation of HS equations, where
these algebraic structures are realized in a manifest way. More
precisely, in addition to the 1-form connection of the
embedding algebra, there is a set of $0$-form fields which are associated to the
basis elements of the dynamical symmetry algebra. The constraints
on these fields are simply the (anti)commutation relations of the algebra.

In this approach all the ingredients needed to formulate HS equations are the algebra of dynamical symmetries and the
embedding algebra. More generally, this gives a new class of gauge invariant equations defined by a pair of superalgebras
so that known HS theories form a particular subclass. Furthermore, we demonstrate that the system admits a
concise AKSZ formulation in terms of the Chevalley-Eilenberg cohomology differentials of the two superalgebras.

Depending on the realization of the HS algebra the equations may describe off-shell theory
in which case the system can be put on-shell through a version of the factorization procedure from~\cite{Vasiliev:2003ev}.
We cover the variety of models including Vasiliev's HS theories in various dimensions as well as variations of the
$d$-dimensional HS theory involving partially-massless or massive fields. The latter two are AdS/CFT dual to non-unitary singletons
and generalized free fields respectively. Finally, we also discuss conformal and massless off-shell fields.

\section{Uniform representation}
\label{sec:uniform}

\subsection{Equations of motion and gauge symmetries}

The equations of motion are built out of the following data: a Lie
(super)algebra $\algg$, which is the algebra of {\em dynamical symmetries} and the associative (super)algebra $\algA$, which is the {\em embedding algebra}.  In the known examples $\algg$ is $u(1)$, $sp(2)$ and $osp(1|2)$ or its extension.
It is convenient to pick basis elements $e_a$ in $\algg$ so that the graded commutation relations are
$\commut{e_a}{e_b}=\cc\fud{c}{ab}e_c$ and the (anti)symmetry
relations are
$\commut{e_a}{e_b}=-(-1)^{\p{e_a}\p{e_b}}\commut{e_b}{e_a}$, where
$\p{e_a}$ is the Grassmann degree of $e_a$. $\algA$ is an embedding
(usually star-product) algebra for a HS algebra under consideration and the product in $\algA$ is denoted by $\star$. In particular, $\algA$ typically contains a Lie subalgebra isomorphic
to $so(d,2)$, \textit{i.e.} $AdS_{d+1}$ spacetime isometries.

Given a space-time manifold with coordinates $x^\mm$, where $\mm=0,\ldots,d$, the fields are
1-form $W=W_\mm(x)\, dx^\mm$ and 0-forms $T_a=T_a(x)$, all taking values in $\algA$. The full set of
equations is
\begin{align}
dW+W\star W&=0\notag\,,\\
dT_a+ \qcommut{W}{T_a}&=0\label{VasSystem}\,,\\
\qcommut{T_a}{T_b}^\pm - \cc\fud{c}{ab} T_c &=0\,,\notag
\end{align}
where $\qcommut{T_a}{T_b}^\pm =T_a \star T_b - (-1)^{\p{e_a}\p{e_b}} T_b\star T_a$. The above system is invariant under the gauge
transformations of the form
\begin{align}
\delta W&=d\xi +\qcommut{W}{\xi}\,, & \delta T_a&= \qcommut{T_a}{\xi}\label{VasAD}\,,
\end{align}
where $\xi=\xi(x)$ also takes values in $\mathfrak{A}$.

Disregarding $x$-dependence, fields $T_a$ can be seen as
components of a map $\tau:\algg \to \algA$  with respect to the
basis $e_a$ so that $T_a=\tau(e_a)$. Then the last equation
in~\eqref{VasSystem} simply implies that $\tau$ is compatible with
the algebra, \textit{i.e.}, $\tau$ is a homomorphism. The Grassmann
degree in $\algg$ and $\algA$ determines the parity of the
component fields. More precisely, if $E_A$ denote basis in $\algA$
and component fields are introduced through $T_a=T^A_a E_A$ and
$W=dx^{\mm} W_{\mm}^A E_A$ then
$\p{T^A_a(x)}=\p{e_a}+\p{E_A}$ and
$\p{W^A_\mm}=\p{E_A}$. Note that one can consistently put to zero
all the fermionic component fields. This bosonic truncation
corresponds to $\tau$ being a homomorphism of superalgebras.

System \eqref{VasSystem} determines a background independent
field theory in the sense that the equations do not involve any
background fields. Moreover, the space-time manifold plays a
passive role in the formulation and can be, at least formally,
taken arbitrary.
However,
background fields enter through the choice of the vacuum solution whose existence
in general restricts the space-time geometry.
In what follows we assume that a fixed vacuum
solution $W=W^0,\, T_a=T_a^0$ is given  and a system is
interpreted as a perturbative expansion around $W^0,T^0_a$. In
HS theories $T^0_a$ are space-time independent, \textit{i.e.} $d
T^0_a=0$, while   $W^0$ is a flat connection of the anti-de Sitter
algebra $so(d,2) \subset \algA$.

Given a vacuum solution global symmetries are by definition gauge transformations that leave the vacuum solution intact,
\begin{equation}
\label{vacuumsymmetries}
 d\xi_0 +\qcommut{W^0}{\xi_0}=0\,, \qquad\qquad\qquad  \qcommut{T^0_a}{\xi_0}=0\,.
\end{equation}
The first equation uniquely fixes the space-time dependence of global symmetry parameters $\xi_0$, while the
second one implies
that $\xi_0$ is $\algg$-invariant in $\algA$. At least locally in space-time
this means that global symmetries are one-to-one with subalgebra $\algB\subset \algA$ of $\algg$-invariants. As we are going to see, in the specific HS theories considered later $\algB$ is the respective HS algebra.

It is worth mentioning that the choice of the vacuum is closely
related to the precise definition of $\algA$. Indeed, two distinct
vacua can either be equivalent being related by a gauge transformation
or nonequivalent depending on particular functional class
used to define a star-product in $\algA$.
Not to mention that the entire physical content of the HS theory
critically depends on the choice of the functional class.

Although system~\eqref{VasSystem} looks very natural and
similar  systems are known in lower dimensions,
in the HS theory context this was first considered
in~\cite{Grigoriev:2012xg}, where it was shown to describe
off-shell constraints and gauge symmetries for HS fields on $AdS$
for $\algg=sp(2)$ and suitable choice of algebra $\algA$ and the
vacuum. Also,  the simplest version with
$\algg=u(1)$ has been proposed in the  earlier
work~\cite{Vasiliev:2005zu} to describe HS fields at the
off-shell level. It is important to note, however, that now we are
mainly concerned with  Vasiliev equations in various dimensions,
where the representation~\eqref{VasSystem} is implicit in the
literature, and where $\algA$ and $\algg$ originate from the
conventional formulation of HS theory~\cite{Vasiliev:1990en,Vasiliev:1992av,
Vasiliev:1992ix, Prokushkin:1998vn, Vasiliev:2003ev}.

Depending on the realization of $\algA$, the equations \eqref{VasSystem},
\eqref{VasAD} may describe off-shell or reducible system. In this
case, in addition to the equations of motion one needs to define a
consistent factorization needed to eliminate unphysical
components. The factorization is determined by an ideal
$\algh\subset \algg$ whose associated fields \{$T_\alpha\}\subset
\{ T_a\}$ are generators of extra gauge symmetry $\delta
T_a=\lambda^\alpha_a \star T_\alpha$ such that the
system~\eqref{VasSystem} is well defined on equivalence classes.
Details of this procedure are explained in Section
~\bref{sec:factor}, where we discuss  $d$-dimensional
higher-spin  equations.

\subsection{Nontriviality and cohomology}
\label{sec:cohomology}
It is instructive to show the way system \eqref{VasSystem} can yield propagating degrees of freedom.
To this end, let us consider its linearization. Denoting the perturbations by $w$ and $t_a$, the linearized system reads
\begin{equation}
\label{VasSystemL}
\begin{aligned}
D_0 w&=0\,, & \qquad\qquad &\delta w=D_0\xi\;,\\
D_0t_a+ \qcommut{w}{T^0_a}&=0\,,\\
\qcommut{t_a}{T^0_b}^\pm+\qcommut{T^0_a}{t_b}^\pm - \cc\fud{c}{ab} t_c &=0\,, &\qquad\qquad &\delta t_a=\qcommut{T^0_a}{\xi}\;,
\end{aligned}
\end{equation}
where $D_0=d+\qcommut{W^0}{\cdot}^\pm$ is the background covariant derivative, $D_0^2 = 0$.

The last line represents  a standard cohomology problem. Indeed, $t_a$ are maps $\mathfrak{g}\rightarrow\algA$ or, equivalently, elements of $\Lambda^1(\mathfrak{g}^*)\otimes \algA$, which by definition is the space of $1$-cochains. From this perspective the last line contains
the cocycle and the coboundary conditions for $t_a$. If, for instance, $\algg$ is such that $\mathbb{H}^1(\mathfrak{g},\algA)$ is empty, then
one can use the gauge symmetry to set $t_a=0$ so that we are left with
\begin{equation}
D_0w=0\,, \qquad\qquad \qcommut{w}{T^0_a}=0\,.
\end{equation}
This system does not contain 0-forms and hence does not describe local degrees of freedom. More precisely, taking into account residual gauge symmetry
and making use of natural technical assumptions, the space of inequivalent solutions is empty. Indeed, $w$ is just a linearized flat connection.

We now turn to the case where $\mathbb{H}^1(\mathfrak{g},\algA)$ is not empty and show that the linearized system
directly leads to the familiar unfolded representation of the linearized HS equations \cite{Vasiliev:1988xc, Vasiliev:1988sa}
(the form is also known as ``on-mass-shell theorem'' in the literature, \cite{Vasiliev:2001wa,Vasiliev:1988sa}). Let $r_a$ parameterize representatives of $\mathbb{H}^1(\mathfrak{g},\algA)$. This means that using gauge symmetry we can replace $t_a$ by $r_a$ so that the system reads
\begin{equation}
\label{inter}
 D_0w=0\,, \qquad\qquad D_0r_a=ad_a\, w\,,
\end{equation}
where $ad_a=\qcommut{T^0_a}{\bullet}$. The second equation can be solved for $w$ modulo elements of $\mathbb{H}^0(\mathfrak{g},\algA)$: $w=\omega + ad_a^{-1} D_0r^{\vphantom{-1}}_a$ and $ad_a\,\omega=0$. Note that $\omega$ can be seen as taking values in $\mathbb{H}^0(\mathfrak{g},\algA)$ which in turn is the HS algebra $\algB$ (cf. \eqref{vacuumsymmetries}).
Plugging this into the first equation of the system \eqref{VasSystemL} we find
\begin{align}
\label{omst}
D_0\omega&=-D_0\, ad_a^{-1} D_0r^{\vphantom{-1}}_a \,,
\end{align}
which has the structure of unfolded HS equations if one identifies $\omega$ as HS connection 1-form and 0-forms $r_a$ as parameterizing curvatures.
This equation merely expresses the linearized curvature of $\omega$ in terms of generalized Weyl tensors (or Riemann tensors at the off-shell level) ensuring that
the system may describe local degrees of freedom. In other words, it specifies a deviation from being a flat connection. However, to see what the dynamical content exactly is the general considerations are not enough and one is to study concrete choice for $\algg,\algA$, and the vacuum solution.

The embedding algebra $\algA$ is a key element of the whole
construction. It should be large  enough to contain the
higher-spin algebra $\algB$ and the algebra $\mathfrak{g}$ of
dynamical symmetries, actually the full image of $U(\mathfrak{g})$. That the dynamics should be nontrivial is a crucial
restriction, which is manifested by nonzero r.h.s. of
\eqref{omst}. In particular, it follows that the algebra $\algA$ cannot be just  a
tensor product $\algB\otimes U(\mathfrak{g})$, for which the two
factors commute. Indeed, in this case $D_0 ad_a^{-1}  = ad_a^{-1} D_0$ since  operators associated
to the commuting subalgebras also commute. Then, using $D_0^2=0$ one finds out that
the right-hand-side of the relation \eqref{omst} vanishes identically.
A reasonable way out is provided by the
twisted product of associative algebras  \cite{TwistedProd}.  It seems to be the most
general construction that allows one to build an algebra that contains two given algebras
as subalgebras but their images do not necessarily commute to each other.
In practice, the twisted product is realized through a specific star-product (see Section \bref{sec:VasilievProduct}).

The following comments are in order.

(i) The linear equations of the form~\eqref{VasSystemL} naturally
appear in describing various HS fields at the free level in the
so-called parent approach. In particular, for $\algg=u(1)$ and
$\algg=sp(2)$ and suitable choice of $\algA$ the
system~\eqref{VasSystemL} describes free HS fields on respectively
flat~\cite{Barnich:2004cr} and AdS
space-time~\cite{Barnich:2006pc} (see
also~\cite{Vasiliev:2005zu,Grigoriev:2006tt,Grigoriev:2012xg}).
Furthermore, equations of motion for nearly generic (including
mixed-symmetry, partially-massless, etc.) HS fields can naturally
be formulated in the form~\eqref{VasSystemL}~\cite{Alkalaev:2008gi,Alkalaev:2009vm,Alkalaev:2011zv}
or~\eqref{omst}~\cite{Skvortsov:2008vs,Boulanger:2008up,Boulanger:2008kw,Skvortsov:2009zu,Skvortsov:2009nv},
which for certain class of fields requires an extension of \eqref{VasSystem} with higher degree forms,
the modification that we do not discuss in detail.

(ii) The system \eqref{VasSystemL} as well as its equivalent
reduction to \eqref{omst} have a simple homological
interpretation. Namely, if one combines both $w$ and $t_a$ into a
homological complex with the differential being $Q=D_0+\Delta$,
where $\Delta$ is the Lie algebra differential of $\mathfrak{g}$, the
system~\eqref{VasSystemL} takes the form $Q\Psi=0$, $\delta
\Psi=Q\Xi$. The differential of the form $D_0 +D_0\, ad_a^{-1} D_0$ in~\eqref{omst} as defined on
$\mathbb{H}(\mathfrak{g},\algA)$-valued fields is just the
differential induced by $Q$ in the cohomology of $\Delta$ (In this form the homological technique for elimination of auxiliary and Stueckelberg fields
was developed in~\cite{Barnich:2004cr,Barnich:2006pc}. Note also a related $\sigma_-$-cohomology
approach of~\cite{Shaynkman:2000ts}).


\section{Specialization to the Vasiliev equations}

At present, there exist several higher-spin theories in various
dimensions, viz., 2d higher-spin theory of matter fields interacting via topological HS
fields \cite{Vasiliev:1995sv}; the minimal 3d HS theory with massless matter fields  \cite{Vasiliev:1992ix};
Prokushkin-Vasiliev $3d$ theory that
admits a one-parameter family of vacua with massive excitations  \cite{Prokushkin:1998bq};
$4d$ bosonic higher-spin theory \cite{Vasiliev:1990cm, Vasiliev:1990en} and its supersymmetric extensions
with any $\cal N$ \cite{Konshtein:1988yg,Konstein:1989ij, Sezgin:2012ag}; $d$-dimensional bosonic system \cite{Vasiliev:2003ev}. All of them can
be further extended by adding internal (Yang-Mills) symmetries, while certain truncations of the spectra are
also possible (see \cite{Vasiliev:1999ba, Sezgin:2012ag} for review). There are also topological higher-spin systems in three \cite{Blencowe:1988gj} and two dimensions \cite{Alkalaev:2013fsa,Alkalaev:2014qpa}
which are HS extensions of Chern-Simons and Jackiw-Teitelboim gravity models.
A brief review  of the Vasiliev equations in $d=2,3,4$ and any $d$ is given  in Appendix
\bref{app:VasilievDictionary}.

%

In this section we show that the system \eqref{VasSystem} reduces to
the Vasiliev equations provided the appropriate choice of the Lie superalgebra
$\algg$. We observe that for all $d\geq 3$ HS systems a Lie superalgebra  $\algg$ contains
$osp(1|2)$ subalgebra.  More precisely, in the Vasiliev system $\algg$-relations are encoded in terms
of the polynomial Serre type relations imposed on a subset of $\algg$-generators.

\subsection{Serre type relations for osp(1,2)}
\label{sec:HSTDynamicsOSP}

Let us first show how the Serre type realization works in the case
of $sp(2)$ algebra and then consider its supersymmetric extension.
Indeed,  the conventional definition of $sp(2)$ algebra relies on
the three commutation relations among three basis elements,
\begin{align}
[H,E]&=+2E\;, & [H,F]&=-2F\;, & [E,F]&=H\;,
\end{align}
where  $[\cdot, \cdot]$ is the Lie product.
Treating the last  relation as a definition we reduce them to two cubic Serre type  relations,
\begin{align}
[[E,F],E]&=+2E\;, & [[E,F],F]&=-2F \;.
\end{align}
Note that there are two independent equations imposed on two basis elements of the Lie algebra.

In the  $osp(1|2)$ superalgebra case
three even basis elements $E,F,H$ are now supplemented with
two odd ones $e,f$. Their non-vanishing  graded commutation relations take the form
\besubeqs
\begin{align}
\{e,e\}&=-2E \;, &\{f,f\}&=2F \;, & \{e,f\}&=H \;,\label{ospdef}\\
[H,e]&=e \;, &[H,f]&=-f \;, &[E,f]&=e \;,& [F,e]&=f \;,\label{osprel}\\
[H,E]&=+2E \;, & [H,F]&=-2F \;, & [E,F]&=H \;,\label{ospconseq}
\end{align}
\esubeqs
where $[\cdot, \cdot]$ and $\{\cdot, \cdot\}$ define  the graded  Lie product.
Relations between odd basis elements  in the first line can be considered as definitions of even
basis elements, while the second line contains nontrivial cubic relations between odd elements.
Relations  in the third line are algebraic consequences of the first two lines.
In this way,  we arrive at four cubic relations between two odd elements
\begin{align}
[\{e,f\},e]&=e\;, &[\{e,f\},f]&=-f\;, &[\{e,e\},f]&=-2e\;, & [\{f,f\},e]&=2f\;, \label{ospfourrel}
\end{align}
which define the $osp(1|2)$ superalgebra. However,
the above Serre type relations can be reduced even further.
Assuming that $\algg$ is embedded into its universal enveloping
algebra $U(\algg)$, and expanding the (anti)commutators we see that
the first two relations are equivalent to the last two relations.

While everything above is true for any Lie (super)algebra with appropriate modifications, the
following property is special for $osp(1|2)$.  We define an even element $\Upsilon \in U(osp(1|2))$,
\begin{align}
\label{Ups}
\Upsilon=[e,f]+\frac12 \;,
\end{align}
which has a rather special property to (anti)commute with (odd)even elements
\besubeqs
\begin{align}
\{\Upsilon, e\}&=0 \;, & \{\Upsilon, f\}&=0 \;,\\
[\Upsilon, E]&=0 \;, & [\Upsilon, F]&=0 \;, & [\Upsilon, H]&=0 \;.
\end{align}
\esubeqs
Obviously, $\Upsilon$  squared  is the quadratic Casimir of $osp(1|2)$ superalgebra, \textit{i.e.},
$\mathbb{C}_2 = \Upsilon^2$. Using  $\Upsilon$ one can show that defining relations of $osp(1|2)$ superalgebra,
namely  the first two of \eqref{ospfourrel} follow from
\begin{align}
\{\Upsilon, e\}&=0\;, & \{\Upsilon, f\}&=0\;,\label{ospminset}
\end{align}
by simply plugging the definitions \eqref{ospdef} into \eqref{osprel}. Two equations \eqref{ospminset} can be rewritten as
\begin{align}
[e^2,f]&=-e\;, & [f^2,e]&=f\;. \label{osptwo}
\end{align}
Again, just as in the $sp(2)$ case, there are two cubic equations  for two basis elements --- five basis elements
of $osp(1|2)$ can be reduced to only two that obey \eqref{ospminset}. The associative algebra
generated by $e,f$ subjected to the above relations is isomorphic to $U(osp(1|2))$  \cite{Bergshoeff:1991dz}.

\subsection{HS equations in three dimensions}
\label{sec:3d}

In the $3d$ HS theory  \cite{Vasiliev:1992ix,Prokushkin:1998bq}, the Lie superalgebra of dynamical symmetries is chosen to be
\be
\algg = osp(1|2)\;.
\ee
Recalling that fields $T_a$ determine a map $\tau:\algg \to \algA$, we unify the images of odd elements
$\tau(e)$ and $\tau(f)$ into a doublet  $S_\ga$, where $\alpha = 1,2$ and the images $\tau(E),\tau(F),\tau(H)$ into a symmetric tensor $T_{\ga\gb}=T_{\gb\ga}$. The $osp(1|2)$ graded relations
\eqref{ospdef} - \eqref{ospconseq}
now read as
\be
[T_{\alpha\beta}, T_{\gamma\rho}] = \epsilon_{\alpha\gamma} T_{\beta\rho} + \text{3 terms}\;,
\qquad
[T_{\alpha\beta}, S_\gamma] = \epsilon_{\alpha\gamma} S_{\beta} + \epsilon_{\beta\gamma} S_{\alpha}\;,
\qquad
\frac{i}{4}\{S_\alpha, S_\beta\} = T_{\alpha\beta}\;,
\ee
where $\epsilon^{\alpha\beta} = - \epsilon^{\beta\alpha}$ is $sp(2)$ invariant form.
The $sp(2)$ indices are raised and lowered as
$S_\alpha = S^\beta \epsilon_{\beta\alpha } $ and $S^\alpha = \epsilon^{\alpha\beta}S_\beta$. In particular,
$\epsilon^{\alpha\beta}\epsilon_{\alpha\gamma}=\delta^\beta_\gamma$.
The factor $\frac{i}4$ is introduced anticipating the star-product realization of $\algA$.

The  Serre type relations  \eqref{osptwo}
uniquely determine $\tau$ and hence all $T_a$ so that the system \eqref{VasSystem} can equivalently be rewritten
in terms of 1-form fields $W$ and $0$-form fields $S_\alpha$ as follows
\besubeqs\label{VasSystemS}
\begin{align}
dW+W\star W&=0\;, \label{VasSystemSA}\\
dS_\ga+[W,S_\ga]_\star&=0\;,\label{VasSystemSB}\\
\frac{i}{4}\, S_\beta \star S_\ga\star S^\beta&=S_\ga\;.\label{VasSystemSC}
\end{align}
\esubeqs
Here, a component form of the last equation~\eqref{VasSystemSC} reproduces relations \eqref{osptwo}
upon  rescaling.

In the Vasiliev theory, the image $\tau(\Upsilon)$ of the element \eqref{ospdef} also denoted as
$\Upsilon$ has been introduced as a new but not independent field.
Taking into account that \eqref{osptwo} can be equivalently represented as \eqref{Ups} and \eqref{ospminset}
the constraint \eqref{VasSystemSC} can be split into the following two equations
\begin{align}
\label{VasSystemSBB}
[S_\ga, S_\gb]_\star&=-2i\,\epsilon_{\ga\gb}(1+\Upsilon)\;, & \{S_\ga,\Upsilon\}_\star &=0\,,
\end{align}
which have a more familiar form of the deformed oscillator
algebra  \cite{Vasiliev:1989qh}. In order to match Vasiliev equations the field $\Upsilon$ needs to be redefined
as $\Upsilon  = B \star \varkappa$, where $B$ is an arbitrary field, while $\varkappa$ satisfying $\varkappa^2=1$,
called Klein operator, is a specific element of the embedding
algebra $\mathfrak{A}$. In our formulation, the Klein operator  is introduced for convenience
in the process of solving equations \eqref{VasSystemS} over
a specific vacuum, see Section~\bref{sec:HSTDynamicsDetails}.
For a suitable choice of $\algA$ the system \eqref{VasSystemS} is exactly the 3d
Vasiliev system \cite{Vasiliev:1992ix,Prokushkin:1998bq} (see also Appendix \bref{app:VasilievDictionary}).
To be precise, one usually adds the covariant constancy $d\Upsilon+[W,\Upsilon]_\star=0$, which is a
consequence of \eqref{VasSystemSB} and \eqref{VasSystemSBB}.

For systems in four  and any dimensions establishing relation with
the known Vasiliev systems requires extra steps, which can be traced back to specific realizations
of the HS algebras. Furthermore,
these Vasiliev systems involve in addition factorization/extra
constraints needed to describe irreducible systems. All these
subtleties are analyzed in some more details in
Sections \bref{sec:gen-d} and~\bref{sec:4dd}.


\subsection{HS equations in any dimensions}
\label{sec:gen-d}

\subsubsection{Off-shell system}\label{sec:off-shell}

The Vasiliev system in $(d+1)$-dimensions is formulated in two
steps: first one formulates the off-shell system and then performs
the consistent factorization which puts the system
on-shell~\cite{Vasiliev:2003ev} (see also
review~\cite{Bekaert:2005vh}). The off-shell system can be
reformulated in the form \eqref{VasSystem}. As an algebra $\algg$
one takes a semi-direct sum
\be
\label{semidirect}
\algg = sp(2) \inplus  osp(1|2)\;.
\ee
If $S_\alpha, T_{\alpha\beta}$ denote fields associated
to $osp(1|2)$ basis elements and $F_{\alpha\beta}$ to $sp(2)$ ones
the last equation in \eqref{VasSystem} reads as
\begin{equation}
\begin{gathered}
\label{1ext}
 \qcommut{F_{\alpha\beta}}{F_{\gamma\delta}}=\epsilon_{\beta\gamma} F_{\alpha\delta}+\ldots\,,\qquad \qcommut{F_{\alpha\beta}}{T_{\gamma\delta}}=\epsilon_{\beta\gamma} T_{\alpha\delta}+\ldots\,,\qquad \qcommut{F_{\alpha\beta}}{S_\gamma}=\epsilon_{\alpha\gamma} S_\beta+\ldots\,,\\
 \frac{i}4\scommut{S_\alpha}{S_\beta}_\star=T_{\alpha\beta}\,, \qquad \qcommut{T_{\alpha\beta}}{S_\gamma}=\epsilon_{\alpha\gamma} S_\beta+\ldots\,,\qquad  \qcommut{T_{\alpha\beta}}{T_{\gamma\delta}}=\epsilon_{\beta\gamma} T_{\alpha\delta}+\ldots\,,
 \end{gathered}
\end{equation} where the ellipsis denote proper symmetrizations.  The first line contains $sp(2)$-relations and $sp(2)$-module structure of $osp(1|2)$ while the second line contains $osp(1|2)$ relations.

Let $F^{0}_{\alpha\beta}$ denote specific vacuum values of
$F_{\alpha\beta}$. This implies that $F^{0}_{\alpha\beta}$ form
$sp(2)$ which makes $\algA$ into an $sp(2)$-module. If in the
above system one puts $F_{\alpha\beta}$ to its vacuum value
$F^0_{\alpha\beta}$ by hand (and hence $F_{\alpha\beta}$ are not
anymore on equal footing with $S,T,W$ fields) it is not
difficult to see that the system coincides with the Vasiliev
system~\cite{Vasiliev:2003ev} (see
Appendix~\bref{app:VasilievDictionary}) provided one redefines  $\Upsilon=B\star\varkappa$ and
reformulates the $osp(1|2)$ relations solely in terms of $S_{\alpha}$
and $B\star\varkappa$ just like in the $d=3$ case presented in Section
\bref{sec:HSTDynamicsOSP}.

It turns out, however, that there is no need to put
$F_{\alpha\beta}=F^0_{\alpha\beta}$ by hand. Indeed, consider  the first equation in~\eqref{1ext} around the vacuum
solution $F_{\alpha\beta}=F^0_{\alpha\beta}$. Then the
cohomological argument given in Section~\bref{sec:cohomology}
applies because $sp(2)$ is simple and the corresponding cohomology is  empty.
It follows that  $F_{\alpha\beta}$ can
be set to its vacuum value $F^0_{\alpha\beta}$ (at least
perturbatively). In so doing one also restricts gauge parameters
to preserve this on-shell gauge choice:
$\qcommut{F^0_{\alpha\beta}}{\xi}=0$. At the same time equations
$dF_{{\alpha\beta}}+\qcommut{W}{F_{\alpha\beta}}=0$ in this gauge
imply $\qcommut{F^0_{\alpha\beta}}{W}=0$ so that both the gauge
parameter and the connection belong to the off-shell HS algebra.
To conclude, the system \eqref{VasSystem} for the constraint algebra given by \eqref{semidirect} yields  a more general theory than the original Vasiliev equations, but they are perturbatively equivalent over the specific vacuum $F_{\alpha\beta}=F^0_{\alpha\beta}$.

The above argument is based on the Whitehead lemma which, strictly speaking, only applies to cohomology with coefficients in
finite-dimensional modules. This is enough if, for instance, the $sp(2)$-cohomology differential preserves the decomposition of $\algA$ into a direct sum of finite-dimensional subspaces. It turns out that this is indeed the case if one takes a standard vacuum solution $F^0_{\alpha\beta}$ for $F_{\alpha\beta}$ because $\qcommut{F^0_{\alpha\beta}}{\cdot}$ as well as the cohomology differential is of vanishing homogeneity in all the oscillators and hence preserve the subspaces of definite homogeneity. This is shown in Section~\bref{sec:embeddingALG}, where explicit definitions for $\algA$ and $F^0_{\alpha\beta}$ are also given.
An example where such a decomposition does not exist and $sp(2)$ relations do have nontrivial solutions can be found in Section
\bref{sec:parent}

\subsubsection{Factorization}
\label{sec:factor}

Now we elaborate on the consistent factorization needed to put the
off-shell system on-shell. We show that factorization can be
performed at the level of the system determined by~\eqref{1ext}.
Moreover, it can be seen as a certain gauge symmetry at the price
of introducing extra fields. This gives a better understanding of
the factorization procedure even in the conventional
formulation  of the Vasiliev system.

In this context it is often convenient to use the following set of
fields~\footnote{The similar trick is used within the Vasiliev
equations, when enforcing $sp(2)$ invariance at the non-linear level \cite{Bekaert:2005vh}} $S_\alpha,T_{\alpha\beta},\bar F_{\alpha\beta}$, where
$\bar F_{\alpha\beta}=F_{\alpha\beta}-T_{\alpha\beta}$. In terms of
the new fields, relations \eqref{1ext} involving $\bar
F_{\alpha\beta}$ take the form
\begin{equation}
\label{1ext1}
 \qcommut{\bar F_{\alpha\beta}}{\bar F_{\gamma\delta}}=\epsilon_{\beta\gamma} \bar F_{\alpha\delta}+\ldots\,,\qquad \qcommut{\bar F_{\alpha\beta}}{S_\gamma}=0\,,\qquad
 \qcommut{\bar F_{\alpha\beta}}{T_{\gamma\delta}}=0 \,,
 \end{equation}
while the relations between $S_\alpha,T_{\alpha\beta}$ stay unchanged.

In terms of the off-shell system the factorization means to eliminate the ideal generated by $\bar F_{\alpha\beta}$. It turns out that system \eqref{VasSystem}
is well-defined on equivalence classes of fields with respect to the following equivalence relation
\begin{equation}
\label{gs-ext-fact}
W\sim W+\lambda^i \star \bar F_i\,, \qquad  T_a \sim T_a +\lambda^j_a \star \bar F_j\,, \quad
\end{equation}
which we present in the infinitesimal form and where $T_a=\{S_\alpha, T_{\alpha\beta}, F_{\alpha\beta}\}$ and
$\bar F_l=\{\bar F_{\alpha\beta}\}$. Equations \eqref{VasSystem} understood as those on equivalence classes can be explicitly written as
\begin{equation}
\label{VasSystem-ext-fact}
\begin{aligned}
dW+W\star W&=u^l \star \bar F_l\,,\\
dT_a+ \qcommut{W}{T_a}&=u^l_a \star \bar F_l\,,\\
\qcommut{T_a}{T_b}^\pm - \cc\fud{c}{ab} T_c &=u_{ab}^l \star \bar F_l\,,
\end{aligned}
\end{equation}
where $u$-fields are not treated as dynamical (in other words the equations only imply that such $u$ do exist). Then the equivalence relations~\eqref{gs-ext-fact} can be seen as the gauge transformations of $W,T_a$. In these terms the consistency of the factorization is just the invariance of the above equations
under~\eqref{gs-ext-fact} supplemented by appropriate transformations of $u$-fields. What one actually checks is that variation of the equations under~\eqref{gs-ext-fact}
is proportional to $\bar F_i$.

The above construction applies to a generic system~\eqref{VasSystem} provided the factorization is performed with respect to the generators of the ideal $\algh\subset \algg$.
In the case at hand, ideal $\algh$ is the diagonal $sp(2)$ in $sp(2)\inplus osp(1|2)$, \textit{i.e.} we have the following
coset
\begin{align}
\frac{sp(2)\inplus osp(1|2)}{sp(2)}\;.
\end{align}
More generally, one can consider consistent factorizations based on
ideals in the enveloping algebra $U(\algg)$ (see below).

Consider now the linearized system by taking
$W=W^0+w,S_\alpha=S^0_\alpha+s_\alpha,
T_{\alpha\beta}=T^0_{\alpha\beta}+t_{\alpha\beta},
F_{\alpha\beta}=F^0_{\alpha\beta}+f_{\alpha\beta}$.
Linearization of the  gauge symmetry~\eqref{gs-ext-fact} allows to gauge away
components of $w,s_\alpha,t_{\alpha\beta},f_{\alpha\beta}$ proportional to $\bar F^0_{\alpha\beta}$. Upon
the elimination of $f_{\alpha\beta},t_{\alpha\beta}$ (using gauge invariance and equations of motion) the system
becomes equivalent to the linearized on-shell Vasiliev system describing massless fields of all integer spins.

One can go even further and consider $u$-fields entering~\eqref{VasSystem-ext-fact} at the equal footing with $W,T_a$. In this interpretation in addition to
gauge transformations of $u$ induced by~\eqref{gs-ext-fact} extra gauge symmetry may be needed to ensure that $u$ do not bring new degrees of freedom.

The core of the above system is the last equation in~~\eqref{VasSystem-ext-fact}. Its gauge symmetry is given by
$\delta T_a=\lambda_a^i \star  \bar F_i+\qcommut{T_a}{\xi}$, where the standard gauge symmetry with parameter $\xi$ has been reinstated and the corresponding gauge transformations for $u$ are assumed. These can be written as follows
\begin{equation}
\label{const-sys-eq}
\begin{gathered}
\qcommut{T_a}{T_a}^\pm=U^c_{ab}\star T_c\,,\qquad \delta T_a=\lambda_a^b \star T_b + \qcommut{T_a}{\xi}\,,\\
\delta U_{ab}^c=\lambda_a^d\star U_{db}^c+\qcommut{T_a}{\lambda_b^c}
-(-1)^{\p{a}\p{b}}(\lambda_b^d\star U_{da}^c+\qcommut{T_b}{\lambda_a^c})+U_{ab}^d\star \lambda_d^c+\qcommut{U_{ab}^c}{\xi}\;,
\end{gathered}
\end{equation}
where $U^c_{ab}$ is to be identified as $\cc^c_{ab}+u^c_{ab}$ and some of the components in $\lambda^a_b$ and $u^c_{ab}$ vanish identically. More precisely,
only those corresponding to the ideal $\algh\subset\algg$ are nonvanishing. Equations~\eqref{const-sys-eq} and gauge symmetries are precisely those defining the constrained Hamiltonian system with constraints $T_a$ and structure functions $U^c_{ab}$ . In particular, gauge transformation with parameter $\lambda^a_b$ is nothing but the infinitesimal redefinition of the constraints
which is a natural equivalence of constrained systems. Further details on the field-theoretical interpretation of constrained Hamiltonian systems can be found in, \textit{e.g.},~\cite{Grigoriev:2006tt,Bekaert:2013zya}. Note, however, that in contrast to the constrained system where all $\lambda^a_b$ can be nonvanishing in our case $\lambda^a_b=0$ if $e_a$ is not in
$\algh$. In other words, the equations for a usual constrained system correspond to $\algh=\algg$.

To complete the discussion of totally-symmetric HS fields in $(d+1)$-dimensions let us mention that the
off-shell system based on $\algg$ \eqref{semidirect} may describe other  on-shell systems if one allows for
more general factorizations. For instance, consider an ideal generated by $\cF_0=C_2-(\lambda^2-1)$, where
$C_2=-\half \bar F_{\alpha\beta}\star\bar  F^{\alpha\beta}$ is the $sp(2)$ Casimir element.
When $\lambda=l$ is an integer there appears an additional ideal, which can be seen to be generated by
\begin{equation}
\label{PM-ideal}
         \cF_{\alpha_1\ldots \alpha_{2\ell}}=\bar F_{(\alpha_1\alpha_2} \star \ldots \star \bar F_{\alpha_{2\ell-1} \alpha_{2\ell})}\;,
\end{equation}
for $l\geq 1$ and where all the $sp(2)$-indices in the
second expression are symmetrized.
Replacing $\bar F_l$ with $\cF_0$ and $\cF_{\alpha_1\ldots \alpha_{2\ell}}$ in \eqref{VasSystem-ext-fact}
(and hence replacing $u^l_{\cdot}$ with $u^{\alpha_1\ldots \alpha_{2\ell}}_{\cdot}$ and $u^0_\cdot$ as well) one
ends up with a consistent system. It is clear that for $\ell=1$ one
recovers \eqref{VasSystem-ext-fact} describing massless fields. The resulting HS system  describes an interacting
system of (partially)-massless fields of depth $t=1,3,\ldots,
2\ell-1$ on $AdS_{d+1}$. The generalization of the Vasiliev theory to partially-massless
fields was suggested in~\cite{Bekaert:2013zya,Alkalaev:2014qpa,Vasiliev-unpublished}.

In order to see partially-massless fields in the spectrum  let us
consider  1-form gauge fields subjected to $sp(2)$ singlet
condition $\qcommut{\bar F_{\alpha\beta}}{W}=0$. These can be decomposed
in $\bar F_{\alpha\beta}$ so that expansion coefficients are identified with
(partially)-massless fields of odd depth~\cite{Skvortsov:2006at,Alkalaev:2007bq}. At the same time
the factorization eliminates the coefficients associated to the ideal in the algebra of $sp(2)$-singlets
 generated by
$C_2-(\lambda^2-1)$ and $\cF_{\alpha_1\ldots \alpha_{2\ell}}$ \cite{Alkalaev:2014qpa} so that
only fields of depth $1,3,\ldots,2\ell-1$ remain.

Let us say a few words about the $AdS/CFT$ interpretation of various
HS theories \cite{Sezgin:2002rt, Klebanov:2002ja}. The bosonic
Vasiliev theory should be generically dual to a free boson theory
in $d$ dimensions \cite{Didenko:2012vh}. For $\lambda=l$ the dual
theory is $\Box^\ell \phi=0$ or multi-critical points of vector
models~\cite{Bekaert:2013zya}.  At generic $\lambda$  we
find massive HS fields in the spectrum, the CFT dual should be a
mean field theory -- a generalized free field of dimension $\frac{d}2\pm \lambda$, depending on the boundary conditions
imposed.

One way to see that massive or partially-massless fields  appear in the spectrum is to make use of the linear relation between
quadratic Casimir elements of the Howe dual algebras $o(d,2)$ and $sp(2)$. Indeed, in our setting $c_2 =  - \frac{1}{4}(d^2-4)  + C_2$,  where $c_2$ and $C_2$ are respectively orthogonal and symplectic Casimir operators \cite{Vasiliev:2004cm}. For $c_2$  one then
finds $c_2 = - \Delta_\lambda(\Delta_\lambda -d)$, where $\Delta_\lambda=\frac{d}{2}-\lambda$.  The irreducible conformal
module with this value is $D(\frac{d}{2}-\lambda,0)$. The spectrum of the respective HS theory in the bulk is
$D(\frac{d}{2}-\ell,0) \otimes  D(\frac{d}{2}-\ell,0)$ which decomposes into irreducible modules of particular
$AdS$ fields.  Note that possible energy  values of these fields are determined by $\lambda$.
For $\lambda=\ell$ the module is the (higher order) singleton $D(\frac{d}{2}-\ell,0)$ whose square
decomposes into (partially)-massless fields of depth $1,3,\ldots,2\ell-1$ (for $\ell=1$ this is the well-known Flato-Fronsdal theorem, the case of $\ell>1$ was in~\cite{Bekaert:2013zya}). For $\lambda$ generic, modules with the non-special values of energy (these are associated to massive fields) appear in the tensor square. Note also that in this case $D(\frac{d}{2}-\lambda,0)$ is a Verma module and hence the conformal scalar it describes is off-shell (from the Verma module perspective equations are associated to singular vectors which are not present as $D(\frac{d}{2}-\lambda,0)$ is irreducible, see e.g.~\cite{Shaynkman:2004vu}).

There is a group theoretical explanation for the ideals $\cF_0=C_2-(\lambda^2-1)$ and \eqref{PM-ideal}. We have an image of $U(\algg)$ in $\algA$ generated by $T_a$ and can consider a more general quotient
\begin{align}\label{Ucoset}
\frac{U\big(sp(2)\inplus osp(1|2)\big)}{C_2-(\lambda^2-1)} ={gl_\lambda}\otimes U(osp(1|2))\;.
\end{align}
The first factor is the Feigin's $gl_\lambda$ \cite{Feigin},
which is defined as a quotient of $U(sp(2))$ by a two-sided ideal generated by $\cF_0$. At generic $\lambda$ algebra $gl_\lambda$ is infinite-dimensional and simple. When $\lambda=l$ is an integer the value of the Casimir is that of the $l$-dimensional irreducible representation of $sp(2)$. In this case $gl_{l}$ contains a two-sided ideal generated by \eqref{PM-ideal} with the quotient being $gl(l)$.


\subsection{HS equations in four dimensions}
\label{sec:4dd}
In the $4d$ case the Lie superalgebra of dynamical symmetries is given by a direct sum
\be
\label{ospdirectsum}
\algg = osp(1|2)\oplus osp(1|2)\;,
\ee
where each factor can be defined along the lines of the previous section using
$sp(2)$ vectors  $S_\ga$ and $\bar S_\gad$. Namely, using  \eqref{VasSystemSBB} one finds
\besubeqs
\begin{align}
\label{4dconstr}
&[S_\ga,S_\gb]_\star= -2i \, \epsilon_{\ga\gb}(1+\Upsilon)\;, &&
[\bar{S}_\gad,\bar{S}_\gbd]_\star= +2i \,\epsilon_{\gad\gbd}(1+\bar{\Upsilon})\;,\\
& \{S_\ga,\Upsilon\}_\star=0 \;,&&\{\bar{S}_\gad,\bar{\Upsilon}\}_\star=0\;, \\
& \hspace{60mm} \{S_\ga,\bar{S}_\gad\}_\star=0\;,
\end{align}
\esubeqs
where $\Upsilon$ and $\bar{\Upsilon}$ are elements \eqref{Ups} associated to two copies of $osp(1|2)$.
The last condition accounts for a direct sum of  Lie superalgebras $osp(1|2)$.

In the standard realization of $\algA$ (see Section \bref{sec:embeddingALG}), this
system does not describe an irreducible HS theory.  The problem is that $\Upsilon$
and $\bar{\Upsilon}$ are related to the selfdual and
anti-selfdual components of the Weyl tensor and its higher-spin
generalizations and hence cannot be independent. It follows that some further constraints that belong to $U(\mathfrak{g})$ are needed.

In order to get an irreducible system we need to take the elementary extension of the Lie superalgebra $\algg$ \eqref{ospdirectsum}.
It means adding  an odd element $K$
that (anti)commutes with  $osp(1|2)\oplus osp(1|2)$ basis elements. In particular, we have
\begin{align}
\{K,S_\ga\}_\star&=0\;, &\{K,\bar S_\gad\}_\star&=0\;, & [K,\Upsilon]_\star&=0\;, & [K,\bar{\Upsilon}]_\star&=0\;.
\end{align}
The additional restriction that makes the system irreducible reads
\begin{align}\label{UglyExtraConstr}
\Upsilon=\bar{\Upsilon} \star K\;.
\end{align}
We can think of $U(osp(1|2))$ as of a noncommutative 2-sphere
$\mathbb{S}_{R}$, whose radius squared is given by the Casimir operator
$R^2=\mathbb{C}_2=\Upsilon^2$. The additional constraint implies $\Upsilon^2=\bar{\Upsilon}^2$, i.e. it is
a square root of $R^2=\bar{R}^2$, so that we have $\mathbb{S}_{R}\times
\mathbb{S}_{R}$.

To get a system that is explicitly equivalent to the original Vasiliev equations one redefines
$\bar{S}_\gad\rightarrow \bar{S}_\gad  \star K$ provided that $K$ is invertible.
It follows that the system \eqref{4dconstr} remains
mainly intact while the only changes  are that
$ \{S_\ga,\bar{S}_\gad\}_\star=0$   goes into $ [S_\ga,\bar{S}_\gad]_\star=0$, and
$[\bar{S}_\gad,\bar{S}_\gbd]_\star$ is now opposite in sign. In the Vasiliev system element $K$ is identified with
the so-called total Klein operator, see Appendix \bref{app:VasilievDictionary}. The supersymmetric extensions
of the $4d$ equations follow the same logic, but we do not discuss them here.

The $4d$ theory must be isomorphic to the $d$-dimensional theory, discussed in the previous section, when $d=4$.
However, they are realized differently. This difference can be attributed to the fact that there are two
realizations of the same HS algebra available in $4d$, which we discuss in Section \bref{HSDynamics}.

\section{HS dynamics and star-product}
\label{HSDynamics}

Here we describe a standard oscillator realization of  the embedding algebra $\algA$ and
respective HS algebra $\algB$ in various  spacetime dimensions. Also,
we review  relevant  types of the star-products  with particular emphasis to the so-called twisted
star-product. It will be shown that the choice of the star-product and/or specific vacuum solution
essentially determines whether respective HS theory has local degrees of freedom or not.

\subsection{Higher-spin algebras }
\label{sec:HSalg}

In Section \bref{sec:uniform} the  HS algebra $\algB$ was introduced
as the symmetry algebra of the vacuum. Indeed, according to the defining relation
\eqref{vacuumsymmetries}  HS algebra is spanned by $\algg$-invariants. Equivalently,
its elements are given by the cohomology  $\mathbb{H}^0(\mathfrak{g}, \algA)$. Obviously,
being a global symmetry algebra,  $\algB$  plays
a fundamental role in HS theory in contrast to the embedding algebra $\algA$ which
is basically  a convenient tool for generating interaction vertices.

Apart from the above definition of $\algB$, there are more physical, but equivalent, definitions that naturally reflect various aspects of HS field dynamic. Basically, this happens because of the AdS/CFT correspondence that identifies gauge symmetries of the bulk theory with the global symmetries of its boundary dual. Below  we briefly consider a few relevant realizations.

\begin{itemize}

\item From the CFT perspective, algebra $\algB$ is the associative algebra of global symmetries of a
conformally invariant field equation. In the simplest case, one
considers the massless  Klein-Gordon equation in $\mathbb{R}^{d-1,1}$ spacetime,
\textit{i.e.}, $\square \phi=0$. It is known to be conformally invariant,
enjoying the symmetry generated by $so(d,2)$ conformal Killing
vectors, which are first order differential operators. Since the
equation is linear one can multiply symmetries to generate an
infinite-dimensional symmetry algebra formed by the differential
operators associated to conformal Killing tensors. It turns out that this algebra exhaust all
global symmetries of $\square \phi=0$ and hence is $\algB$ itself~\cite{Eastwood:2002su}.
The  corresponding Noether currents  are generated  by
totally-symmetric conserved tensors
\be
\label{cc}
j_{\mu_1...\mu_s}= \pl_{(\mu_1} \cdots \pl_{\mu_k}\phi^*\; \pl_{\mu_{k+1}} \cdots \pl_{\mu_s)} \phi - \text{traces}\;,
\quad \qquad \pl^{\mu_1}j_{\mu_1...\mu_s}=0\;,
\ee
where $s=1,2,..., \infty$ \cite{Konstein:2000bi}.

\item Algebra $\algB$  can also be thought of as  a unique algebra
generated by the stress tensor $T_{\mu\nu}$ and at least
one higher-spin conserved tensor $j_{\mu_1...\mu_s}$, $s>2$ of  some CFT \cite{Fradkin:1986ka, Maldacena:2011jn, Boulanger:2013zza, Stanev:2013qra, Alba:2013yda}.
The uniqueness theorem holds under further assumptions
that the only conserved tensors are totally symmetric ones and
there is a special case of $4d$ where one finds a one-parameter
family, \cite{Fradkin:1989md, Fernando:2009fq, Boulanger:2011se,
Boulanger:2013zza, Govil:2013uta,Manvelyan:2013oua}, provided the locality is relaxed.

\item From the AdS perspective, algebra  $\algB$ is the global symmetry algebra
of a given HS theory at the linearized level,  or, equivalently, a symmetry algebra of the
most symmetric vacuum. It can also be seen as a gauge algebra of HS theory in the sense that the 1-form HS connections take values in $\algB$, \cite{Vasiliev:1980as, Vasiliev:1986td, Lopatin:1988hz,
Vasiliev:2001wa}. One should stress, however, that the algebra structure of $\algB$ is deformed at the nonlinear level.

\item One can formalize above realizations by taking the universal enveloping algebra
$U(so(d,2))$ and factor out a two-sided ideal $Ann(S)$ that
annihilates the irreducible $so(d,2)$-module $S$ \cite{Fradkin:1990ir,Eastwood:2002su},
\be
\label{HSquot}
\algB = U(so(d,2))/Ann(S)\;.
\ee
The module $S$ is a spin-$0$ Dirac singleton representation spanned by
solutions of $\square \phi=0$.  The other way around, the symmetry algebra of $S$ is the endomorphism algebra of $S$, \textit{i.e.},
$S\otimes S^*$, which coincides with $U(so(d,2))/Ann(S)$.

\end{itemize}

\subsection{Oscillator realization}
\label{sec:oscillator}

For applications, an efficient realization of HS algebras is needed. It turns out that in all known cases
bosonic HS algebra can be realized as  the Weyl algebra (or its quotient) which
is the associative algebra generated by operators with canonical commutation relations.

In particular, it implies that AdS algebra $so(d,2) \subset \algB$ enjoys an oscillator realization.
Indeed, a simple but crucial fact is that given a set of $n$
variables  $\zeta^\aA$ satisfying
\be
\label{canonical}
[\zeta_\aA,\zeta_\aB]_\star =2i\mathcal{C}_{\aA\aB}\;,
\ee
their quadratic combinations form $sp(2n)$ algebra with
$\mathcal{C}^{\aA\aB}$ being the invariant  tensor,
\begin{align}
&[T_{\aA \aB},T_{\aC\aD}]_\star=\mathcal{C}_{\aA\aC}T_{\aB\aD}+ \text{3 terms}\;,
& T_{\aA\aB}&=\frac{1}{4i} \{\zeta_\aA,\zeta_\aB\}_\star\,.
\end{align}
The symplectic algebra is either isomorphic  to AdS
algebra in lower dimensions or contains AdS algebra as a
subalgebra in higher dimensions as listed in the table below.
\be
\notag
\begin{array}{|c|c|c|c|c|}
\hline
\mbox{dim} & AdS-\mbox{algebra}    & \mbox{isomorphism}                      & \mbox{Lorentz-algebra} & \mbox{isomorphism} \\ \hline
  2 & so(1,2) & sp(2,\mathbb{R})                        & so(1,1) & u(1)       \\
  3 & so(2,2) & sp(2,\mathbb{R})\oplus sp(2,\mathbb{R}) & so(2,1) & sp(2,\mathbb{R})        \\
  4 & so(3,2) & sp(4,\mathbb{R})                        & so(3,1) & sp(2,\mathbb{C})_{\mathbb{R}}\\
  d+1 & so(d,2)\subset sp(2d+4) &            & so(d,1)\subset sp(2d+2) & \\ \hline
\end{array}
\ee
The oscillator realizations of the AdS algebra together with its splitting into Lorentz subalgebra and
translations are summarized in the table below
(see, \textit{e.g.}, reviews \cite{Vasiliev:1999ba,Bekaert:2005vh}).
\footnote{The $sp(2)$ indices are $\ga,\gb,...=1,2$ and $\gad,\gbd,...=1,2$; the $sp(4)$
indices $\aAs,\aBs,...=1,...,4$ can be split into a pair of $sp(2)$ ones, $\aAs=(\ga,\gad)$; the AdS $so(d,2)$
indices are $A,B,...=0,...,d+1$, the Lorentz $o(d,1)$ indices are
$a,b,...=0,...,d$. Tensors $\eta^{AB}$, $C_{\aAs \aBs}$ and $\epsilon^{\ga\gb}$ are the invariant  metrics   of
$so(d,2)$, $sp(4)$ and $sp(2)$, respectively.}

\be\notag
\begin{array}{|x{1cm}|c|c|c|c|}
\hline
  \mbox{dim}   &  \mbox{generators} &    AdS &   \mbox{Lorentz} &   \mbox{Translations} \tabularnewline\hline
  2\rule{0pt}{54pt}     &  [y_\ga,y_\gb]_\star=2i\epsilon_{\ga\gb} & T_{\ga\gb}=\frac{1}{4i}\{y_\ga, y_\gb\}_\star &
  L=\frac{1}{4i}\{y_1, y_2\}_\star & \subsB{P_1=\frac{1}{2i} y_1 y_1}{P_2=\frac{1}{2i}y_2 y_2} \tabularnewline\hline
  3\rule{0pt}{36pt}     &  \subsB{[y_\ga,y_\gb]_\star=2i\epsilon_{\ga\gb}}{\psi^2=1} & L_{\ga\gb},  \quad P_{\ga\gb}
  & L_{\ga\gb}=\frac{1}{4i}\{y_\ga, y_\gb\}_\star & P_{\ga\gb}=\psi L_{\ga\gb} \tabularnewline\hline
  3\rule{0pt}{54pt}     &  \subsC{[y_\ga,y_\gb]_\star=2i\epsilon_{\ga\gb}(1+\nu k)}{\{y_\ga,k\}_\star=0}{\psi^2=1, k^2=1} & L_{\ga\gb},  \quad P_{\ga\gb}
  & L_{\ga\gb}=\frac{1}{4i}\{y_\ga, y_\gb\}_\star  & P_{\ga\gb}=\psi L_{\ga\gb} \tabularnewline\hline
  4\rule{0pt}{54pt}  & \subsB{[y_\aAs,y_\aBs]_\star=2iC_{\aAs \aBs}}{y_\aAs=(y_\ga, \bar{y}_\gad)}& T_{\aAs\aBs}=\frac{1}{4i}\{y_\aAs,y_\aBs\}_\star & \subsB{L_{\ga\gb}=\frac{1}{4i}\{y_\ga, y_\gb\}_\star }{L_{\gad\gbd}=\frac{1}{4i}\{\bar{y}_\gad, \bar{y}_\gbd\}_\star } &  P_{\ga\gad}=\frac{1}{2i}y_\ga \bar{y}_\gad  \tabularnewline\hline
  d+1\rule{0pt}{36pt}  & \subsB{[Y^A_\ga,Y^B_\gb]_\star=2i\eta^{AB}\epsilon_{\ga\gb}}{Y^A_\ga=(y^a_\ga, y^{\phantom{a}}_\ga)} & T^{AB}=\frac{1}{4i}\{Y^A_\ga,Y^{B\ga}\}_\star & L^{ab}=\frac{1}{4i}\{y^a_\ga, y^{b\ga}\}_\star & P^a=\frac{1}{2i}y^a_\ga y^{\ga}\tabularnewline\hline
\end{array}
\ee

\vspace{5mm}

Note that  the  $3d$ AdS algebra is a direct sum of two $sp(2)$.
This doubling is achieved using  an additional element $\psi$, $\psi^2=1$, which makes Clifford algebra in one dimension.
Also, both two and three dimensional AdS algebra generators can be built  via the so-called deformed oscillators
with commutation relations parameterized by a continuous  $\nu$ \cite{Wigner,Vasiliev:1989qh,Vasiliev:1989re},
see the third line in the table.

According to \eqref{HSquot}, the HS algebra is defined as a
quotient. In lower dimensions, using $sp(2n)$ oscillators allows
to resolve the ideal (see, \textit{e.g.}, \cite{Gun}). It follows that in $d=2,3,4$ dimensions  HS
algebras are identified with the enveloping algebra of the
relations in the table above, \textit{i.e.} an element of the HS
algebra is a function $f(y)$ or $f(y,k)$ in two dimensions, $f(y,\psi)$ or $f(y,\psi,k)$
in three dimensions, and $f(y,\bar{y})$ in four dimensions, see, \textit{e.g.} \cite{Vasiliev:1999ba} for review.

It is worth noting that without the $\psi$ element in $3d$ case, the
enveloping algebra of the deformed commutation relations is isomorphic to $U(osp(1|2))/I$, where $I$ is the two-sided ideal
generated by the shifted Casimir element $(C_2-\frac14(\nu^2-1))$ \cite{Bergshoeff:1991dz}. The subalgebra
of even in $y$ elements  decomposes into a direct sum of two $gl_\lambda$ (which was
defined after \eqref{Ucoset}) for $C_2=\frac14(\nu^2\pm3\nu-3)$.

In $d+1$ dimension, the AdS algebra is only a subalgebra of $sp(2d+4)$, so that the ideal is
only partially resolved and certain further constraints are needed. The oscillator approach
developed in \cite{Vasiliev:2003ev} makes  the Weyl algebra generated by oscillators $Y^A_\alpha$
a bimodule of two algebras $so(d,2)-sp(2)$, where $\tau_{\ga\gb}=-\frac{i}4\{Y^A_\ga,Y^B_\gb\}\eta_{AB}$
are the $sp(2)$ generators, which form a Howe dual pair, \textit{i.e.} $[T^{AB},\tau_{\ga\gb}]=0$. Then, $\algB$ can be realized
as an $sp(2)$-invariant subspace of the Weyl algebra further quotiented
by a two-sided ideal spanned by the elements proportional to $sp(2)$ generators
\begin{align}
\algB \ni f (Y^A_\ga)&: && [f(Y), \tau_{\ga\gb}]_\star=0\;, && f(Y)\sim f(Y)+ g^{\ga\gb}(Y)\star \tau_{\ga\gb}\;.
\end{align}
The Taylor coefficients
of the quotient representatives  $f(Y)$  carry $so(d,2)$ indices described
by traceless two-row rectangular Young diagrams of arbitrary length
\begin{align}
f(Y^A_\ga)= \sum_k f_{A_1...A_k, B_1...B_k} T^{A_1B_1}... \;T^{A_kB_k}\;, && f_{A_1...A_k, B_1...B_k}=
\parbox{50pt}{\begin{picture}(50,20)\linethickness{0.4mm}\put(0,0){\line(1,0){50}}\put(0,0){\line(0,1){20}}
\put(0,10){\line(1,0){50}}\put(10,0){\line(0,1){20}}\put(40,0){\line(0,1){20}}\put(50,0){\line(0,1){20}}
\put(0,20){\line(1,0){50}}\put(20,12){\footnotesize{$k$}}\end{picture}}\;\;\;.
\end{align}
Naturally, these tensor expansion coefficients are
in one-to-one correspondence with both connections of the HS algebra $\algB$,  and   conformal
Killing tensors discussed  in Section  \bref{sec:HSalg}.

Let us mention again that considering more general ideals one can
get HS algebras for partially-massless and massive fields following the same
procedure as in Section \bref{sec:factor} with $\bar{F}_{\ga\gb}$ replaced by
$\tau_{\ga\gb}$. The CFT interpretation is that these are the algebras of higher symmetries of the
polywave equation\footnote{As it was already mentioned, such algebras involve partially-massless fields of even depth only. Denoting the solution space of order-$2l$ polywave equation as $S_l$ these algebras are isomorphic to $S_l\otimes S^*_l$. In order to get algebras for partially-massless fields of odd depth one needs to construct a matrix-type extension whose entries are $S_{l}\otimes S^*_{l'}$ for some $l$, $l'$. Then the fields of odd depth live in off-diagonal blocks.} $\Box^\ell \phi=0$ \cite{Bekaert:2013zya}, when $\lambda=l$ is an integer and the algebra of symmetries of generalized free field of dimension $\frac{d}2\pm\lambda$.

We stress that the
$2d$, $3d$ (at $\nu = 0$) and $4d$ algebras are isomorphic to
the $d$-dimensional algebra for $d=2,3,4$ provided that the
functions of respectively $y_\ga$, $y_\ga$, and $y_\aAs$,  are
restricted to be even, \textit{i.e.}, half-integer spins are projected out\footnote{A relation between vector and spinor realizations of $d=2,3,4$ HS algebras is explicitly
discussed in  \cite{Vasiliev:2004cm}. }.
Each of the oscillator realizations given above has its own
features that do not bear any invariant meaning in contrast to the
HS algebra itself. However, these features affect the choice of
$\mathfrak{g}$ and, hence, the realization of the embedding
algebra $\algA$.

\subsection{Twisted star-product}
\label{sec:VasilievProduct}

As we argued in the introduction the embedding algebra $\algA$ has the structure of a twisted product, where it is the twist that is responsible for nontriviality of the theory. The factors are the higher-spin algebra $\algB$ and the full algebra of dynamical symmetries, i.e. the image of $U(\mathfrak{g})$ in $\algA$. This is the structure present in all HS theories constructed so far.
Because HS algebras admit oscillator realizations it is possible to give a concrete
realization of the twisted product as the star-product \cite{Vasiliev:1990vu}. Below we
collect some basic definitions and properties of star-products.

The star-product algebra is the algebra of functions in commuting variables $\zeta^\aA$ that is equipped with a non-commutative product, called star-product,
\begin{align}\label{generalStarProduct}
(f\star g)(\zeta)=f(\zeta)\exp{i\left(\frac{\overleftarrow{\pl}}{\pl \zeta^\aA}\, \Omega^{\aA\aB} \frac{\overrightarrow{\pl}}{\pl \zeta^\aB}\right)} g(\zeta)\;.
\end{align}
The anti-symmetric component of $\Omega^{\aA\aB}$, $\mathcal{C}=\frac12(\Omega-\Omega^T)$ is the symplectic metric,
which specifies the commutation relations $[\zeta^\aA,\zeta^\aB]_\star=2i\mathcal{C}^{\aA\aB}$.
The symmetric part of $\Omega^{\aA\aB}$ is responsible for ordering prescription for the operators that $\zeta^A$ are symbols of. For example, the matrix $\Omega^{\aA\aB}$ that corresponds to the totally-symmetric ordering is just $\mathcal{C}^{\aA\aB}$:
\begin{align}
\label{symmetric}
\;\;\;\mbox{symmetric}&: &&(f\star g)(\zeta)=f(\zeta)\exp{i\left(\frac{\overleftarrow{\pl}}{\pl \zeta^\aA}\, C^{\aA\aB} \frac{\overrightarrow{\pl}}{\pl \zeta^\aB}\right)} g(\zeta)\;.\qquad\qquad\qquad\qquad\qquad\qquad
\end{align}
Another commonly used  prescription is the normal  ordering, which corresponds to particular
splitting of $\zeta^\aA$ into $q^m$ and $p_n$, \textit{i.e.} $\zeta^\aA=(q^m,p_n)$. Then, the product
implementing the $qp$-ordering is
\begin{align}
\mbox{normal}&: &&(f\star g)(q,p)=f(q,p)\exp{i\left(\frac{\overleftarrow{\pl}}{\pl p_n} \frac{\overrightarrow{\pl}}{\pl q^n}\right)} g(q,p)\;,\qquad\qquad\Omega=\begin{bmatrix}
                                                                              0 & I \\
                                                                              0 & 0 \\                                                                       \end{bmatrix}\;.
\end{align}

It is remarkable that the higher-spin  theory uses a specific mixture of  normal and symmetric
orderings, which was introduced by Vasiliev  \cite{Vasiliev:1990vu}. We will refer to it as the twisted
star-product. Suppose $\zeta^\aA$ with $\aA = 1,...,4$ splits  into a pair of variables $\zeta^\aA=(y^\ga, z^\gb)$, where $\alpha, \beta = 1,2$. The twisted
star-product corresponds to the symmetric ordering among $y^\ga$ and among $z^\gb$ with the symplectic structure
given by the epsilon-symbol  $\epsilon^{\ga\gb}$ in both the sectors, while it is normal ordered with
respect to $q^\ga=y^\ga+z^\ga$ and $p^\ga=y^\ga-z^\ga$. Namely, let  $\algA_0$ be an algebra of functions in $y^\ga$ and $z^\ga$ endowed with the following product:
\begin{align}\label{twsitedsp}
\phantom{\mbox{twisted}}& &&(f\star g)(y,z)=f(y,z)\exp{i
\left(\frac{\overleftarrow{\pl}}{\pl y^\ga}+\frac{\overleftarrow{\pl}}{\pl z^\ga}\right)\epsilon^{\ga\gb}\left( \frac{\overrightarrow{\pl}}{\pl y^\gb}-\frac{\overrightarrow{\pl}}{\pl z^\gb}\right)} g(y,z)\;.
\end{align}
or, more generally, with a one-parameter family of star-products:
\begin{align}\label{twsitedspB}
\;\;\;\;\mbox{twisted}_\vartheta&: &&(f\star g)(y,z)=f(y,z)\exp{\begin{pmatrix}
                                \frac{\overleftarrow{\pl}}{\pl y^\ga}\,, & \frac{\overleftarrow{\pl}}{\pl z^\ga} \\
                              \end{pmatrix} \begin{pmatrix}
  \epsilon^{\ga\gb} & -\vartheta\epsilon^{\ga\gb} \\
  \vartheta\epsilon^{\ga\gb} & \epsilon^{\ga\gb} \\
\end{pmatrix} \begin{pmatrix}
   \frac{\overrightarrow{\pl}}{\pl y^\gb} \\
   \frac{\overrightarrow{\pl}}{\pl z^\gb} \\
 \end{pmatrix}
 } g(y,z)\;,
\end{align}
interpolating between \eqref{twsitedsp} at $\vartheta=1$ and the symmetric product \eqref{symmetric} at $\vartheta=0$.
Note, that at $\vartheta=0$ there is no mixing among $y^\alpha$ and $z^\beta$.
In Section \bref{sec:HSTDynamicsDetails} we show that for $\vartheta\neq0$ the HS theory is nontrivial,
while taking the limit  $\vartheta=0$ yields a
topological theory.

It is important to stress that all star-products are equivalent
when restricted to polynomials. However, in the HS theory certain non-polynomial elements appear in perturbative solution of
\eqref{VasSystem}. With this being
said, we have to consider $\vartheta\neq0$ twisted   and $\vartheta=0$ symmetric
star-products as non-equivalent. Then, the twisted
star-product can be thought of as a particular example of the general concept of
twisted tensor product of associative algebras.

Indeed, let $A$ and $B$ be two associative algebras. According to Ref.
\cite{TwistedProd}, algebra $C$ is a twisted tensor product of $A$
and $B$ iff there exists injective algebra homomorphisms $i_A:
A\rightarrow C$ and $i_B: B\rightarrow C$ such that the linear map
$i_A\otimes i_B : A\otimes B\rightarrow C$ defined by $(i_A\otimes
i_B) (a\otimes b)=i_A(a) \star i_B(b)$ is a linear isomorphism. Here $a\in A$, $b\in B$ and $\star$ is a product in $C$.

In the case of interest we have the associative higher-spin algebra $\algB$ and the algebra of dynamical symmetries. In all
known cases the nontrivial part of the embedding algebra $\algA$
has the same realization as the twisted star-product
\eqref{twsitedsp}, where $A$ and $B$ are the star-product algebras
of functions $a(y^\ga)$ and $b(z^\ga)$, respectively, with the
products $\mu_A$ and $\mu_B$ given by
\begin{align}
&\mu_A=\exp{+i\frac{\overleftarrow{\pl}}{\pl y^\ga}\epsilon^{\ga\gb}\frac{\overrightarrow{\pl}}{\pl y^\gb}}\;,
&&\mu_B=\exp{-i\frac{\overleftarrow{\pl}}{\pl z^\ga}\epsilon^{\ga\gb}\frac{\overrightarrow{\pl}}{\pl z^\gb}}\;.
\end{align}
The algebra $C$ is the algebra of functions $c(y,z)$ equipped with
the twisted product \eqref{twsitedspB}. The map $\mu_A\otimes \mu_B$
determined by
\begin{align}
a \otimes b\mapsto  i_A(a) \star i_B(b)&= a(y)\tau(\vartheta) b(z)\;,&&\tau(\vartheta)=\exp{-i\vartheta\frac{\overleftarrow{\pl}}{\pl y^\ga}\epsilon^{\ga\gb}\frac{\overrightarrow{\pl}}{\pl z^\gb}}\;,
\end{align}
is an isomorphism for a suitable class of functions because $\tau$ is formally invertible:
$\tau^{-1}(\vartheta)=\tau(-\vartheta)$. At $\vartheta=0$ we get back
to the usual product of associative algebras. It is crucial for nontriviality of the theory
that $[i_A(a(y)),i_B(b(z))]\neq0$ while both $\algB=A$ and $U(\algg)=B$ are subalgebras of $C$.

\subsection{Embedding algebra and vacuum}
\label{sec:embeddingALG}

The section is aimed at defining the embedding algebra $\algA$ and
the vacuum solution for $T_a$ and $W$. The general structure of
the algebra $\algA$ is a twisted product of the HS algebra $\algB$
and the algebra of dynamical symmetries $U(\mathfrak{g})$. Therefore,
$\algA$ always includes the generators/relations we used to define
$\algB$. The twist enters through one or more factors of the
algebra $\algA_0$, which is the twisted star-product algebra generated by
$y_\ga,z_\ga$, where $y_\alpha$'s belong to the realization of $\algB$.
A number of discrete elements, which can be combined into
Clifford algebras, can also appear. It is always possible to take
the tensor product with matrix algebras $Mat_n$, which allow
higher-spin  fields to carry Yang-Mills indices. The
algebras so defined can be truncated by some reality
conditions and other (anti)-automorphisms. For example, the
Yang-Mills factor can be truncated to compact forms $su$,
$so$, $usp$ \cite{Konshtein:1988yg,Konstein:1989ij}.

In the table below we list some of the cases where $\algA$ is known. Below, $A_{d+1}$ denotes the Weyl
algebra formed by $y^a_\ga$ in the $(d+1)$-dimensional HS algebra and the relations determining the
star-product in the sector of $(y_\alpha,z_\alpha)$ and ($y_\gad,z_\gad$) variables are those of $\algA_0$ and are omitted.
\be\notag
\begin{array}{|x{1cm}|c|c|c|} \hline
  \mbox{dim}            &    \mbox{generators, relations}       & \algA   & \mbox{vacuum} \tabularnewline\hline
        3\rule{0pt}{16pt}   & y_\ga,z_\ga,\{\psi_i,\psi_j\}=2\delta_{ij}  &  \algA_0\otimes Cl_{2,0} & S^0_\ga=z_\ga\tabularnewline\hline
    4\rule{0pt}{16pt}   & y_\ga,z_\ga, \,\, \bar{y}_\gad,\bar{z}_\gad & \algA_0\otimes\algA_0 & S^0_\ga=z_\ga,\, S^0_\gad=\bar{z}_\gad \tabularnewline\hline
      d+1\rule{0pt}{44pt}   & \subsB{y_\ga,z_\ga}{[y^a_\ga, y^b_\gb]_\star=2i\eta^{ab}\epsilon_{\ga\gb}} & \algA_0  \otimes A_{d+1} & \subsB{S^0_\ga=z_\ga}{{F}^0_{\ga\gb}=\frac{1}{4i}\big(\{Y^a_\ga,Y^b_\gb\}_\star\eta_{ab}+\{y_\ga, y_\gb\}_\star -\{z_\alpha, z_\beta\}_\star\big)}\tabularnewline\hline
\end{array}
\ee

An essential ingredient of the theory is the vacuum solution $W^0$, $T^0_a$. In HS theories 1-form $W^0$ is a flat
connection of the anti-de Sitter algebra $so(d,2)\subset \algB$.
By definition, the background $W^0$ has vielbein $h^a$ and spin-connection $\varpi^{a,b}$ as its components along
Lorentz and translation generators. For instance, in the $d$-dimensional notation we have
$W^0=\frac12 L_{ab} \varpi^{a,b}+P_ah^a$. Then, any non-degenerate solution of the flatness condition
$dW^0+W^0\star W^0= 0$  \eqref{VasSystem} describes empty anti-de Sitter space.

According to \eqref{vacuumsymmetries}, the HS algebra $\algB$ is the centralizer of the vacuum. Since the
generators $T_a$ of the $osp(1|2)$ part can be reduced to a two-component
field $S_\ga$, it is sufficient to specify a vacuum value only for
$S_\ga$. It is always $S^0_\ga=z_\ga$. It is obviously consistent
with $W_0$ since $[z_\ga,f]_\star=-2i\pl_\ga f$, where
$\pl_\ga=\frac{\pl}{\pl z^\ga}$ and the generators of the HS
algebra are $z$-independent. Therefore,
$dT^0_a+[W^0,T^0_a]_\star=0$ is satisfied. For the same reason,
the global symmetries of the vacuum, which are solved from
\eqref{vacuumsymmetries}, belong to
$\mathbb{H}^0(\mathfrak{g},\algA)$ and form the HS algebra.

In Section~\bref{sec:off-shell} we proved that fluctuations of $F^0_{\alpha\beta}$ are trivial.
The proof crucially relies on the assumption that ad($F^0_{\alpha\beta})=\qcommut{F^0_{\alpha\beta}}{\cdot}$ are of vanishing homogeneity degree in $Y_\alpha^a, y_\ga,z_\alpha$. While it is obviously true for the $y^a_\ga$ part of the generators, its validity for $y_\ga,z_\ga$ relies on an important property of the twisted star-product. Because of the twisting that take place in \eqref{twsitedspB} the action of each of the two $sp(2)$ subalgebras associated with $y_\alpha$ and $z_\alpha$ is deformed, for example,
\begin{align}
\label{adsp2}
\frac12\xi^{\ga\gb}[L^y_{\ga\gb}, f(y,z)]_\star&=\xi^{\ga\gb}\left(y_\ga-i\vartheta\frac{\pl}{\pl z^\ga}\right)\frac{\pl}{\pl y^\gb} f(y,z)\;, && L^y_{\ga\gb}=-\frac{i}4\{y_\ga,y_\gb\}_\star\;,
\end{align}
but the diagonal $sp(2)$ algebra that contributes to $F^0_{\ga\gb}$ still acts canonically
\begin{align}
\label{adsp2d}
\frac12\xi^{\ga\gb}[L^y_{\ga\gb}+L^z_{\ga\gb}, f(y,z)]_\star&=\xi^{\ga\gb}\left(y_\ga\frac{\pl}{\pl y^\gb}+z_\ga\frac{\pl}{\pl z^\gb}\right) f(y,z)\;, && L^z_{\ga\gb}=\frac{i}4\{z_\ga,z_\gb\}_\star\;.
\end{align}
Therefore the Whitehead lemma used in Section~\bref{sec:off-shell} can be applied.

\subsection{Linearized HS dynamics and star-product}
\label{sec:HSTDynamicsDetails}
In what follows, it will be shown that the nontriviality  of a given HS theory depends essentially
on the choice of the star-product. With an appropriate choice of the star-product
we show that the cohomology groups $\mathbb{H}^1(\mathfrak{g},\algA)$ that parameterize the deviation from the flat connection are not empty and \eqref{omst} describes free fields of all spins.

Let us consider first-order perturbations \eqref{VasSystemL} of the system \eqref{VasSystem}.
We expand $W=W^0+w$ and $S_\ga=z_\ga+ s_\ga$. The linearized equations together with the gauge transformations have the form
\footnote{For the $4d$ system there is a doubling $s_\ga$, $s_\gad$, which we do not consider in detail.}
\besubeqs\label{VasLinearized}
\begin{align}
D_0 w&=0\;,& \delta w&=D_0\xi\;,\label{VasLinearizedA}\\
D_0 s_\ga&= \pl_\ga w\;, & &\label{VasLinearizedB}\\
\{z_\ga, \pl_\gc s^\gc\}_\star&=0\;, & \delta s_\ga&=\pl_\ga \xi\;,\label{VasLinearizedC}
\end{align}
\esubeqs
where $D_0$ is the background covariant derivative $D_0=d+[W^0,\cdot]_\star\equiv d+ad_{W^0}$ and $\pl_\gc s^\gc$
is the linearization of $\Upsilon$, see e.g. \eqref{VasSystemSBB}. To get these equations we used
$[z_\ga, \cdot]_\star=-2i\pl_\ga$, where
$\pl_\ga=\frac{\pl}{\pl z^\ga}$.

Equation \eqref{VasLinearizedC} is convenient to solve assuming  that $\algA$ contains an element called Klein operator (see, \textit{e.g.}, \cite{Vasiliev:1999ba}) that implements the $\mathbb{Z}_2$
automorphism,
\begin{align}
\label{klein}
\rho(\zeta^\aA)=\varkappa \star \zeta^\aA \star \varkappa^{-1}=-\zeta^\aA\;.
\end{align}
The choice of ordering prescription affects the functional form of the Klein operator.
For instance, see also Appendix B in \cite{Iazeolla:2011cb} for discussion of different orderings,
\be\label{KleinOperators}
\begin{aligned}
\mbox{symmetric}&: & \varkappa&=\delta(\zeta) & \qquad\qquad\qquad
\mbox{normal}&: & \varkappa&=\exp \left(i q^m p_m\right)\\
\mbox{twisted}&: & \varkappa&=\exp \left(i z_\ga y_\gb \epsilon^{\ga\gb}\right)&
\mbox{twisted}_\vartheta&: & \varkappa&=\exp \left(i\vartheta^{-1} z_\ga y_\gb \epsilon^{\ga\gb}\right)
\end{aligned}
\ee
Ignoring the issue  of functional class we can map anti-commutators to commutators using
the Klein operator. Indeed, according to \eqref{klein}, the vacuum satisfies
 $\varkappa\star z_\ga\star \varkappa^{-1}=-z_\ga$, so that
\begin{align}
\pl_\gc s^\gc \equiv \Upsilon \;:\qquad \quad  \{z_\ga, \Upsilon\}_\star=[z_\ga, \Upsilon\star \varkappa]_\star\star \varkappa^{-1}=0 \qquad \Longleftrightarrow &\qquad  \pl_\ga (\Upsilon\star \varkappa)=0\;,
\end{align}
where the last equation is true thanks to invertibility of $\varkappa$. It implies that $\Upsilon\star\varkappa$
is $z$-independent,
\begin{align}
\label{upss}
\Upsilon \star\varkappa= C(y|x)\;,
\end{align}
where $C(y|x)$ is an arbitrary function of all variables but $z_\ga$,\textit{ i.e.} of
$x^{\underline{m}}$, $y_\alpha$ and possibly of  $y^a_\ga$ or $\psi_i$, depending on the theory we consider. In order to reconstruct the gauge potential $s_\ga$ from $\Upsilon$ we represent  equation $\pl_\gc s^\gc  =  \Upsilon$
in the dualized form as follows
\begin{align}
\pl_\ga s_\gb-\pl_\gb s_\ga=\epsilon_{\ga\gb}\Upsilon=\epsilon_{\ga\gb}C(y|x)\star\varkappa\;.
\end{align}
At this point it is useful to translate everything into the language of differential forms, by contracting all indices with anticommuting differentials $dz^\ga$. Then the last equation is simply
\begin{align}
\pl s&=\Upsilon \, dz^\ga\wedge dz^\gb\epsilon_{\ga\gb}\;,
\end{align}
where $\pl=dz^\ga\pl_\ga$ is the $2d$ de Rham differential. Then, the general solution is
$ s= \pl^{-1}(\Upsilon)+\pl \xi $, where $\pl^{-1}$ is any representative of anti-derivative, and
the last term represents exact forms  \eqref{VasLinearizedC}.
For example, one can use the standard contracting homotopy for the de Rham complex to obtain
\begin{align}
\label{s}
s= \pl^{-1}(\Upsilon)\equiv\pl^{-1}(\Upsilon\, \epsilon_{\ga\gb}dz^\ga\wedge dz^\gb)=z^\ga \int_0^1 t\,dt\,
\epsilon_{\ga\gb}\Upsilon(zt)\, dz^\gb\;,
\end{align}
where we worked in the Schwinger-Fock gauge $z^\ga s_\ga=0$.
We would like to stress that $\Upsilon$ does depend on $z_\alpha$ because of the Klein operator $\varkappa$, cf. \eqref{upss}.

Now we lift expression for $s_\ga$ \eqref{s} to equation \eqref{VasLinearizedB}, which again has a form of $\pl w=D_0 s$ and can be solved as before,
$w(y,z|x)=\omega(y|x)+\pl^{-1}D_0 s(y,z|x)$, where homogeneous part satisfies $ \pl \omega(y|x)=0$.
According to the general discussion of Section \bref{sec:cohomology},
$\omega $ represents cohomology
$\mathbb{H}^0(\mathfrak{g},\algA)$. It follows that the HS field $\omega$ is identified with
$\mathbb{H}^0(\mathfrak{g},\algA)$ connection. Using  identities  $\{d,\pl^{-1}\}\equiv0$ and $\pl^{-1}\pl^{-1}\equiv0$
the solution can be simplified to
$w=\omega+\pl^{-1}\left(ad_{W_0} s\right)$.
On substituting this to the first equation \eqref{VasLinearizedA} we can restrict to $z=0$ surface to get
\begin{align}
D_0\omega&=\left.-ad_{W^0}\,\pl^{-1}ad_{W^0}\,\pl^{-1} \Upsilon\right|_{z=0}  \label{OMST}\;.
\end{align}
Thus, we arrive at the particular realization of \eqref{omst} used in  the Vasiliev theory. The dynamical content of the equation \eqref{OMST} relies on the particular choice
of the star-product. \textit{E.g.}, using the twisted star-product at  $\vartheta\neq0$ in $d$ dimensions
yields the standard unfolded equations
\begin{align}
D_0\omega(y|x)&= h^a\wedge h^b \epsilon_{\ga\gb}\frac{\pl^2}{\pl y^a_\ga \pl y^b_\gb}\,C(y^a_\ga,y_\ga=0|x) \label{OMSTd}\;,
\end{align}
where function $C(y|x)$ on the right-hand-side  parameterizes all spin-$s$ Weyl tensors (see \cite{Bekaert:2005vh} for more details).

Let us consider again equations~\eqref{VasLinearizedC}.
Taking the limit $\vartheta=0$ one finds out that the star-product corresponds to the symmetric
ordering and the Klein operator is realized as the $\delta$-function, \eqref{KleinOperators}.
Even without Klein operator it is obvious that equation $\{z_\ga, \Upsilon\}_\star=2z_\ga \Upsilon=0$ has only a
trivial regular solution $\Upsilon =0$. Therefore, the embedding algebra based on the untwisted product of $\algB$ and $U(\mathfrak{g})$ leads to a topological system for the particular vacuum $S^0_\ga=z_\ga$.

\subsection{Relation to parent system}
\label{sec:parent}

As we have just seen the nontriviality of the Vasiliev system has
to do with the nonvanishing symmetric part \eqref{twsitedspB} of
$\Omega^{\aA\aB}$ in the star-product \eqref{generalStarProduct},
which makes the Klein operator regular. Using the formulation in
$(d+1)$ dimensions we now demonstrate that it is nevertheless
possible to describe degrees
of freedom using just untwisted star-product.
In this case the nontriviality enters through the
specific choice of the vacuum solution.

Having found that $\pl_\nu s^\nu=0$ at $\vartheta=0$ and hence $s_\ga$ is pure gauge we observe that the second equation in \eqref{1ext1} then implies $\dl{z^\nu}\bar F_{\alpha\beta}=0$ and the extended system takes the form
\begin{equation}
\label{sp2par}
 d W + W\star W=0\,, \qquad d  \bar F_{\alpha\beta}+\qcommut{W}{\bar F_{\alpha\beta}}=0\,, \qquad \qcommut{\bar F_{\alpha\beta}}{\bar F_{\gamma\delta}}=\epsilon_{\beta\gamma} \bar F_{\alpha\delta}+\ldots\,,
\end{equation}
known in the literature~\cite{Grigoriev:2012xg}. Note that this is again of the type~\eqref{VasSystem} with $\algg=sp(2)$.

According to the discussion in~\bref{sec:off-shell}, taking as a vacuum solution $\bar F_{\alpha\beta}=\frac{1}{4i}\{Y^A_\alpha, Y^B_{ \beta}\}_\star\eta_{AB}$ allows to perturbatively eliminate $\bar F_{\alpha\beta}$. However, the system
can describe degrees of freedom if the vacuum is chosen differently. Following~\cite{Grigoriev:2012xg} we first fix the allowed class of functions to be polynomials in $Y_2^A$ with coefficients
in formal series $Y^A_1$. With this choice the vacuum
\begin{equation}
 \bar F^0_{\alpha\beta}=\left.\frac{1}{4i}\{Y^A_\alpha, Y^B_{ \beta}\}_\star\eta_{AB}\right|_{Y_1^A\to Y^A_1+V^A}
\end{equation}
is not equivalent to the one without shift in $Y_1$ (the shift is
not well-defined for formal series). As a consequence, the
linearized system is non-empty and was shown
in~\cite{Grigoriev:2012xg} to describe massless fields of all
integer spins at the off-shell level (\textit{i.e.}, equivalent to
the linearized Vasiliev system before factorization). Note that in
this case the argument based on Whitehead lemma does not work
because $\commut{\bar F^0_{\alpha\beta}}{\cdot}$ is not
homogeneous in $Y$ (in particular, $\Delta$-cohomology is nonempty
in degree $1$ and, in a certain sense, is precisely a
configuration space of HS fields).

A closely related system makes sense in the context of conformal HS fields on the $d$-dimensional boundary. More precisely, replacing $AdS_d$ with its boundary and the AdS compensator $V^A$
satisfying $V^AV_A+1=0$ with the conformal one satisfying $V^AV_A=0$ the system \eqref{sp2par} describes, in particular, off-shell conformal HS fields. More precisely, the system
describes totally symmetric conformal HS gauge fields at the off-shell level provided one performs a consistent factorization, as described in \bref{sec:factor},
with respect to the  $sp(2)$ algebra. Remarkably, these fields can be seen as leading boundary values for the bulk HS fields. For more details see~\cite{Bekaert:2013zya}.


\subsection{Vasiliev theory squared}
\label{app:VasilievSquared}

It is instructive to see what happens if instead of the Lie superalgebra $\algg=osp(1|2)$ one takes the  Lie algebra $\algg = sp(2)$ while keeping the algebra $\mathfrak{A}_0$ and the vacuum as in Section~\bref{sec:HSTDynamicsDetails}.
The odd generators $S_\alpha$ of $osp(1|2)$ are now absent and the $sp(2)$ generators $T_{\ga\gb}$ satisfy
\begin{align}
[T_{\ga\gb}, T_{\gc\gd}]_\star&=\epsilon_{\ga\gd}T_{\gb\gc}+ \text{3 terms}\;.
\end{align}
If we take the same vacuum $T^0_{\ga\gb}=\frac{i}4\{z_\ga,z_\gb\}_\star$ as before then the second
equation in \eqref{vacuumsymmetries} is a second order equation because of \eqref{adsp2}. Namely, using \eqref{adsp2} one finds
that global symmetries $\xi$ are solved from
\begin{align}
[T_{\ga\gb}, \xi]_\star&=\frac{i}4\{z_\ga, [z_\gb, \xi]_\star\}_\star+(\ga\leftrightarrow\gb)=\left(z_\ga+i\frac{\pl}{\pl y_\alpha}\right)
\frac{\pl }{\pl z_\beta} \xi+(\ga\leftrightarrow\gb)=0\;,
\end{align}
whose general solution involves two arbitrary functions of $y_\alpha$, namely,
\begin{align}
\xi=\xi_0(y)+\pl^{-1}\left(\{dz^\ga z_\ga, \xi_1(y)\}_\star \star\varkappa\right)\;,
\end{align}

where $\pl=dz^\ga\pl_\ga$.
Therefore, $\xi \in \mathbb{H}^0(\mathfrak{g},\algA)$ turns out to contain two branches. These are parameterized
by original  $\xi_0(y)$ appearing in the $osp(1|2)$ case, and  additional $\xi_1(y)$ in the $sp(2)$ case.

The advantage of having a bosonic oscillator realization of the
superalgebra $osp(1|2)$ is that the vacuum odd generators act
as $[z_\alpha, \cdot]_\star   = -2i\frac{\pl}{\pl z_\ga} $ and the
centralizer of the vacuum, which is the HS algebra, is independent of
the auxiliary variables $z_\ga$. In the $sp(2)$ case the vacuum
bosonic generators $[T_{\alpha\beta}, \cdot]_\star$ are second
order operators giving rise to the centralizer bigger than the
original HS algebra.

When solving the field equations one can either ignore the second
branch  at every order of the perturbative expansion or define a certain projector
that explicitly removes these modes. One way or another, every solution in  the
$osp(1|2)$ case is a solution in the $sp(2)$ case as well.

It is worth noting that the above extra branch arises  due to
the particular choice of the star-product and the vacuum which were previously used in the
standard $osp(1|2)$ case. On the other hand, in Section \bref{sec:parent} we showed that
the extra branch in $\mathbb{H}^0(\mathfrak{g},\algA)$ can be avoided by taking a slightly different vacuum.

\section{Algebraic structure and AKSZ form}
\label{sec:algebra}
In this section we discuss the structure of the basic system \eqref{VasSystem} in some more mathematical details. Starting with a Lie superalgebra $\algg$ and associative superalgebra $\algA$ let us consider the superspace of linear maps $\tau: \algg \to \algA$, where $\algA$ is understood as a Lie superalgebra, i.e. with the Lie operation $\qcommut{f}{g}=f\star g-(-1)^{\p{f}\p{g}}g\star f$. The Grassmann degree on the space of maps originates from those on $\algg$ and $\algA$. More precisely, if $\algg=\algg_{\bar 0}\oplus \algg_{\bar 1}$ and $\algA=\algA_{\bar 0}\oplus \algA_{\bar 1}$ are the decompositions into homogeneous components then degree-$0$ maps sends $\algg_{\bar 0}$ to $\algA_{\bar 0}$ and $\algg_{\bar 1}$ to $\algA_{\bar 1}$ while degree-$1$ map sends $\algg_{\bar 0}$ to $\algA_{\bar 1}$ and $\algg_{\bar 1}$ to $\algA_{\bar 0}$. The condition that $\tau$ is a homomorphism reads as
\begin{equation}
\label{hom}
 \tau(a)\star \tau(b)-(-1)^{\p{a}\p{b}}\tau(b) \star \tau(a)=\tau(\commut{a}{b})\,, \quad \forall a,b\in \algg\,.
\end{equation}
If $e_a$ denote a basis in $\algg$ then the above condition takes the form of the third equation in~\eqref{VasSystem}.

There is a natural equivalence on the superspace of homomorphisms:
\begin{equation}
 \tau\sim  \mathcal{I}_\algA\circ \tau \circ \mathcal{I}_\algg\,,
\end{equation}
where $\mathcal{I}_\algA$ and $\mathcal{I}_\algg$ are inner automorphisms of respectively $\algA$ and $\algg$.
An infinitesimal versions of the above equivalence relations read as~\footnote{Everywhere in this section $\qcommut{A}{B}$ denotes the supercommutator $A\star B-(-1)^{\p{A}\p{B}}B\star A$, where $\p{A}$ stands for the total Grassmann degree of $A$.}
\begin{equation}
\tau(a) \sim \tau(a)+\qcommut{\tau(a)}{\xi}\,, \qquad \tau(a) \sim \tau(a) + \tau(\commut{a}{\beta})\,,
\end{equation}
where $\xi \in \algA$ and $\beta \in \algg$. In the context of HS theories the equivalence $\mathcal{I}_\algA$ is interpreted as a genuine equivalence. Indeed, the above transformation is precisely the gauge symmetry~\eqref{VasAD} in the sector of $T_a$ variables. At the same time $\mathcal{I}_\algg$ is treated as a physical symmetry. Note that interpreting $\mathcal{I}_\algg$ as a gauge symmetry can also be useful though we do not have meaningful examples at the moment.

It turns out that the superspace of homomorphisms subject to the equivalence relation generated by $\mathcal{I}_\algA$ completely determines the gauge invariant system~\eqref{VasSystem}. To see this it is convenient to switch to the language of $Q$-manifolds.
First one defines a supermanifold associated to the superspace. Namely, if $E_A$ is a basis in $\algA$ the components
of the homomorphism are $\tau(e_a)=T^A_a E_A$. One then reinterprets $T^A_a$ as coordinates on a supermanifold $M_0$
by prescribing Grassmann parity as $\p{T^A_a}=\p{E_A}+\p{e_a}$. Note that we assume \eqref{hom} imposed so that $M_0$ is a surface in the space of coordinates $T^A_a$
singled out by $\qcommut{T_a}{T_b}=\cc_{ab}^c T_c$.

In order to take into account the gauge symmetry one promotes parameters $\xi$ to ghost coordinates extending $M_0$ to $M$. More precisely, introducing components of the gauge parameter $\xi$ through $\xi=\xi^A E_A$, $\xi^A$ are promoted to coordinates $W^A$ such that
$\p{W^A}=\p{E_A}+1$ and $\gh{W}=1$; ghost degree of $T^A_a$ is zero. Finally, $M$ is equipped with the odd nilpotent
vector field $Q$ determined by
\begin{equation}
\label{Q-str}
 QT_a=\qcommut{W}{T_a}\,, \qquad
 QW=\half\qcommut{W}{W}\,.
\end{equation}
The gauge symmetry induced by $Q$ is precisely the above equivalence $\mathcal{I}_\algA$.

Given a $Q$-manifold equipped with a nonnegative ghost degree one can define a free differential algebra on a given space-time manifold (see e.g.~\cite{Vasiliev:2005zu,Barnich:2005ru} for more details). Namely, to each coordinate $\psi^I$ of ghost degree $p$ one associates a $p$-form field $\Psi^I$ on the space-time manifold. Furthermore, if $p\geq 1$ then coordinate $\psi^I$ also gives rise to a gauge parameter $\epsilon^I$ which is a $(p-1)$-form. The equations of motion and gauge symmetries read as
\begin{equation}
\label{Q-unf}
d\Psi^I+Q^I(\Psi)=0\,, \qquad \delta_\epsilon \Psi^I=d \epsilon^I-\epsilon^J \ddl{Q^I(\Psi)}{\Psi^J}\,.
\end{equation}
Applying the above construction to the $Q$ manifold $M$ one finds $0$-form fields $T_a^A$ and 1-form fields $W^A=dx^\mu W^A_\mu$ (by slight abuse of notation we use the same symbol for a coordinate on $M$ and its associated field) along with the $0$-form gauge parameters $\xi^A$ associated to $W^A$. Equations and gauge transformations~\eqref{Q-unf} are then equations~\eqref{VasSystem} and \eqref{VasAD}. Note, however, that the third equation in~\eqref{VasAD} does not arise this way
but is satisfied thanks to the definition of $M$.

There is a natural way to encode the third equation from~\eqref{VasSystem} by a certain  extension of
the $Q$-manifold $M$ and by using a more general AKSZ framework. The extension amounts to introducing extra coordinates
of the negative ghost degree needed to incorporate the constraints on $T_a$ into the $Q$-structure. This can be done in a nice way using the BRST machinery. Indeed, introducing ghost variables $c^a$ such that $\gh{c^a}=1$ and $\p{c^a}=\p{e_a}+1$ let us consider polynomials in $c^a$ with values in $\algA$. The coordinates on the extended supermanifold $\manM$ are components of
a generic element of this algebra
\begin{equation}
\label{56}
 \begin{gathered}
\Psi=\sum_{k=0}^\infty \psi^A_{a_1\ldots a_k}c^{a_1}\ldots c^{a_k} \,  E_A\,,\\
\gh{\psi^A_{a_1\ldots a_k}}=\gh{E_A}-k\,,\quad \p{\psi^A_{a_1\ldots a_k}}=\p{E_A}+\p{e_{a_1}}+\ldots+\p{e_{a_k}}-k                                                                                                                                                                                                        \end{gathered}
\end{equation}
so that $\gh{\Psi}=1$ and $\p{\Psi}=1$. Note that coordinates $\psi^A_a$ and $\psi^A$ are precisely $T^A_a$ and $\xi^A$ introduced above. The $Q$-structure on $\manM$ is introduced as follows
\begin{equation}
\label{57}
 Q\Psi=\half\qcommut{\Psi}{\Psi}+q\Psi\,,\qquad q=-\half c^a c^b \cc_{ab}^c \dl{c^c}\,.
\end{equation}
It is easy to check that it coincides with~\eqref{Q-str} for $\psi^A=\epsilon^A$ and $\psi^A_a=T^A_a$ while in general acts nontrivially on $\psi^A_{a_1\ldots a_k}$ with $k\geq 1$. The two terms in $Q$ have a simple interpretation: the second term originates from the cohomology differential of the Lie superalgebra $\algg$ while the first one from that of $\algA$ understood as a Lie superalgebra. The later identification becomes clear if one considers $\psi^A$ as a ghost variable associated
to a basis element $E_A$.

Given a $Q$-manifold equipped with the not necessarily
non-negative ghost degree the AKSZ
procedure~\cite{Alexandrov:1995kv} (for a review and further
details see \textit{e.g.}~\cite{Barnich:2009jy,Grigoriev:2012xg})
determines an associated gauge theory. In this case, in addition
to equations~\eqref{Q-unf} there are extra algebraic equations
associated with coordinates of ghost degree $-1$:
\begin{equation}
 Q^I(\Psi)=0\,, \qquad \gh{\psi^I}=-1\,.
\end{equation}

There are no fields associated to coordinates of negative ghost
degree. In particular, the coordinates with negative degree are
put to zero in $Q^I(\Psi)$ entering the above formula. Coordinates
with ghost degree $-2$ and higher do not produce new equations of
motion. In fact they are needed to encode identities (identities
between identities, etc.) between the equations in the
Batalin-Vilkovisky description of the AKSZ system. To conclude,
the equations of motion and gauge symmetries of the AKSZ system
defined by \eqref{56} and  \eqref{57} are
respectively~\eqref{VasSystem} and \eqref{VasAD}. In the
particular case where $\algg$  is a Lie algebra (not superalgebra)
the above AKSZ system was originally proposed
in~\cite{Grigoriev:2012xg}.

As a final remark let us mention that the consistent factorization
given in Section~\bref{sec:factor} can be also naturally embedded
into the AKSZ framework.  If $c^\alpha$ denote ghost variables
associated to the ideal $\algh\subset \algg$ determining the
factorization then in  addition to ghost variables $c^a$ one
introduces the ghost momenta $b_\alpha$, $\gh{b_\alpha}=-1$
conjugated to $c^\alpha$. By allowing $\Psi$ to depend on
$b_\alpha$ as well this leads to extra fields including, in
particular, $u$-fields in~\eqref{VasSystem-ext-fact}.  The
extended AKSZ system then naturally incorporates equations
\eqref{VasSystem-ext-fact} and gauge transformations
\eqref{gs-ext-fact}.


\section{Conclusions}
\label{sec:Conclusions}
In this paper we  have attempted to uniformize all known Vasiliev higher-spin  theories within a single framework given by system \eqref{VasSystem}, whose algebraic origin is manifest. We observed that the specific features of the realization of HS algebra $\algB$ do affect the choice of dynamical symmetries $\algg$ and the embedding algebra $\algA$. It would be interesting to try to avoid any specific realization of HS algebras and give an invariant description of HS theories.

More generally, system \eqref{VasSystem} provides a class of integrable, as we expect, models that are defined by the following data: (i) the symmetry algebra of the vacuum, which in the HS context is the HS algebra, $\algB$; (ii) the algebra of dynamical symmetries, $\mathfrak{g}$, namely by the image of $U(\mathfrak{g})$ in $\algA$, which in the HS story is related to the Lorentz algebra. For example, in $3d$ and $4d$ theories the image is such that we get the enveloping algebra of
the vacuum Lorentz algebra. In $d$-dimensional theory the relation to the Lorentz algebra is made
somewhat implicit because of the Howe duality, \cite{Bekaert:2005vh}.

With these two data one can construct the embedding algebra $\algA$ as a twisted product of $U(\mathfrak{g})$ and $\algB$ and write \eqref{VasSystem}. The dynamics at the linearized level is determined by connections of $\mathbb{H}^0(\mathfrak{g},\algA)=\algB$ whose curvatures are not zero but given by $\mathbb{H}^1(\mathfrak{g},\algA)$. The dynamics is nontrivial if the r.h.s. of \eqref{omst} is non-vanishing, possibly at higher orders of the perturbation theory.

Any theory is substantially characterized by its observables. A
natural class of observables for \eqref{VasSystem}, advocated in
\cite{Sezgin:2005pv,Sezgin2011, Colombo:2012jx}, is given by Casimir operators of
$\mathfrak{g}$, i.e. invariant polynomials of $T_a$. Another type
of observables are Wilson loops $tr\,\mbox{P}\!\exp \oint W$.
Wilson loops can be generalized to decorated Wilson loops where
there are insertions of any functions of $T_a$ in the adjoint
representation of $\algA$. It was shown in
\cite{Didenko:2012tv} that all correlation
functions in Vasiliev theory can be computed in terms of such
observables. It would be interesting to prove that the models
described by \eqref{VasSystem} are integrable at least for a
subset of observables in the sense of having a free-field
realizations, as it happens for the holographic $S$-matrix in the
HS theories with boundary conditions preserving full HS algebra.

A question, which we leave for further developments, is how big is
the class of higher-spin  theories that are covered by the system
\eqref{VasSystem}. The known HS theories of Vasiliev type do not
exhaust all possible higher-spin  fields. In dimensions higher than
four the spin degrees of freedom, which are characterized by
irreducible (spin)-tensors of the Wigner little group can be of
more general symmetry type than just totally symmetric and the
spectrum of string theory involves such fields.

\section*{Acknowledgements}
\label{sec:Aknowledgements}
We would like to thank Glenn Barnich, Xavier Bekaert, Nicolas Boulanger, Slava Didenko, Carlo Iazeolla,
Per Sundell and Mikhail Vasiliev for many useful discussions. This work was initiated and partially performed during visits of K.A. and M.G. at the
Albert Einstein Institute, Golm.
The work of K.A. was supported by RFBR grant No 14-01-00489. The work of E.S. was supported in part by RFBR grant No 14-02-01172. The work of M.G. was
supported by Russian Science Foundation grant 14-42-00047.


\begin{appendix}
\renewcommand{\thesection}{\Alph{section}}
\renewcommand{\theequation}{\Alph{section}.\arabic{equation}}
\setcounter{equation}{0}\setcounter{section}{0}



\section{Standard form of Vasiliev equations}
\label{app:VasilievDictionary}

\paragraph{2d system.} In $2d$ dimensions we distinguish between two types of higher-spin systems:
first are topological models and second are models with propagating matter fields. Both of them are of the form \eqref{VasSystem} with $\algg=u(1)$. The choice of $\algA$ depends on the presence or absence of local degrees of freedom. It follows that the two-dimensional fields are given by 1-form $W(x)$ and 0-form $T(x)$
subjected to the BF equations of motion
\be
\label{2dim}
\begin{array}{l}
dW+W\star W=0 \;,
\\

dT+ \qcommut{W}{T}=0\;.
\end{array}
\ee
The embedding algebra $\algA$ can be taken either finite-dimensional \cite{Alkalaev:2013fsa} or
infinite-dimensional
\cite{Fradkin:1989uh,Vasiliev:1995sv,Alkalaev:2014qpa}.
The particular choice of infinite-dimensional $\algA$ yields local degrees of freedom \cite{Vasiliev:1995sv}.
In both cases the system \eqref{2dim} follows from $2d$ BF action functional.
\paragraph{$3d$ system.} The full system of equations has the form \cite{Vasiliev:1992ix}
\begin{align}
\label{4d-vas}
d W+W\star W&=0\;, & \{S_\ga, B\star \varkappa\}_\star&=0\;,\\
d(B\star \varkappa)+[W, B\star \varkappa]_\star &=0\;, & [S_\ga, S_\gb]_\star&=-2i\epsilon_{\ga\gb}\left(1+ B\star \varkappa \right)\;,\\
d S_\ga+[W, S_\ga]_\star&=0\;.
\end{align}
Let us note that the Prokushkin-Vasiliev system, \cite{Prokushkin:1998vn}, can be cast into the same form as above, but $\algA$ is slightly different. The bosonic projection is made by the following kinematical constraints
\begin{align}
[\varkappa,B]_\star =0\;, && [\varkappa,W]_\star =0 \;,&& \{\varkappa, S_\ga\}_\star =0\;,
\end{align}
where $\varkappa$ is the Klein operator. Alternatively, this is just a condition that $W$ and $S$ belong to respectively even and odd components of $\algA$ or
in the formulation given in Section~\bref{sec:uniform}, $S_\alpha$ determine a parity-even map from $\algg$ to $\algA$. These constraints imply that $\varkappa$ can be removed from the third equation, giving simply $dB+[W, B]_\star =0$.

\paragraph{$4d$ system.} The full system of equations has the form \cite{Vasiliev:1992av}
\besubeqs\label{VasAll}
\begin{align}
&dW+W\star W=0\,,\\
&d(B\star \varkappa)+[W, B\star \varkappa]_\star=0\,,\\
&dS_{\ga}+[W,S_{\ga}]_\star =0\,,
&&d\bar{S}_{\gad}+[W, \bar{S}_{\gad}]_\star =0\,,\\
&[S_{\ga},
S_{\gb}]_\star =-2i\epsilon_{\ga\gb}(1+B\star \varkappa)\,,
&&[\bar{S}_{\gad},
\bar{S}_{\gbd}]_\star =-2i\epsilon_{\gad\gbd}(1+B\star \bar{\varkappa})\,,\label{BBbar}\\
&\{S_{\ga}, B\star \varkappa\}_{\star }=0\,,
&&\{\bar{S}_{\gad}, B\star \bar{\varkappa}\}_{\star }=0\,,\\
&[S_{\ga}, \bar{S}_{\gad}]_\star =0\,,
\end{align}
\esubeqs
along with kinematical constraints ensuring the theory is bosonic
\begin{align}
\label{4dbosonic}
[K,B]_\star  & =0\;, & [K,W]_\star &=0\;, & \{K, S_\ga\}_\star &=0\;, & \{K,\bar{S}_\gad\}_\star &=0\;,
\end{align}
where $K=\varkappa\star \bar{\varkappa}$ is the total Klein operator, $K\star K=1$. Thanks to the bosonic projection  $\varkappa$ can be replaced with $\bar{\varkappa}$ in the second equation.
Again, conditions~\eqref{4dbosonic} are equivalent to bosonic truncation introduced in Section~\bref{sec:uniform}. Let us note that the extra constraint \eqref{UglyExtraConstr} we had to impose is a way to say that $\Upsilon=B\star\varkappa$ and $\bar{\Upsilon}=B\star\bar{\varkappa}$ from \eqref{BBbar} originate from the same $B$.
\paragraph{$d$-dimensional system.} The full system of equations has the form \cite{Vasiliev:2003ev}
\begin{align}
d W+W\star W&=0 & \{S_\ga, B\star \varkappa\}_\star&=0\;,\\
d(B\star \varkappa)+[W, B\star \varkappa]_\star&=0 \;, & [S_\ga, S_\gb]_\star&=-2i\epsilon_{\ga\gb}\left(1+ B\star \varkappa \right)\;,\\
d S_\ga+[W, S_\ga]_\star&=0\;,
\end{align}
supplemented with the constraints from the $sp(2)$-factor of the coset
\begin{align}
[F^0_{\ga\beta}, W]_\star &=0\;, & [F^0_{\ga\beta}, B]_\star &=0\;, & [F^0_{\ga\beta}, S_\gamma]_\star &=\epsilon_{\ga\gamma}S_\beta + \epsilon_{\beta\gamma}S_\alpha\;,
& [F^0_{\ga\beta}, \varkappa]_\star &=0\;.
\end{align}
Taking into account the explicit realization of $\algA$ and $F^0_{\alpha\beta}$ as a star product algebra the above conditions imply that
bosonic truncation from Section~\bref{sec:uniform} is fulfilled automatically.

Let us stress that in the original paper \cite{Vasiliev:2003ev} the oscillators $Y^A_\ga$ were doubled by introducing $Z^A_\ga$, which have extra components $z^a_\ga$ as compared to $z_\ga$ we used. Correspondingly, there were more fields $S^A_\ga$ introduced. However the equations for the Lorentz components $S^a_\ga$ have the form of Weyl algebra $[S^a_\ga,S^b_\gb]=-2i\epsilon_{\ga\gb}\eta^{ab}$ since there is no deformation due to $B$. Therefore, $S^a_\ga=z^a_\ga$ is an exact solution to all orders and oscillators $z^a_\ga$ can be removed from the definition of $\algA$, as we did, while the field $S^a_\ga$ can be removed from the Vasiliev equations.



\end{appendix}
%

{\footnotesize
\providecommand{\href}[2]{#2}\begingroup\raggedright\endgroup
}

\end{document}